\documentclass[a4paper,11pt]{article}
\usepackage[dvipdfmx]{graphicx}
\usepackage{bm}
\usepackage{braket}
\usepackage{jheppub,ulem} 
\usepackage[T1]{fontenc}

\usepackage{tikz, tikz-feynhand}
\usetikzlibrary{intersections, calc, arrows.meta}

\title{QFT approach to dressed particle processes in preheating and non-perturbative mechanism in kinematically-forbidden regime}

\author[a]{Hidetoshi~Taya}
\author[b]{Yusuke~Yamada}

\affiliation[a]{RIKEN iTHEMS, RIKEN, Wako 351-0198, Japan}
\affiliation[b]{Waseda Institute for Advanced Study, Waseda University, 1-21-1 Nishi Waseda, Shinjuku, Tokyo 169-0051, Japan}

\emailAdd{hidetoshi.taya@riken.jp}
\emailAdd{y-yamada@aoni.waseda.jp}

\preprint{RIKEN-iTHEMS-Report-22}

\abstract{
We provide a quantum-field theoretic formulation of dressed particle dynamics that systematically includes particle production and scattering/decay processes in the preheating era.  Our approach is based on the so-called perturbation theory in the Furry picture, in which coherent background fields (i.e., inflaton and the expanding Universe) are treated non-perturbatively whereas interactions between dressed particles are taken into account perturbatively.  As an application, we consider the instant preheating mechanism and compute the number of produced particles explicitly.  We find a novel non-perturbative particle-production mechanism, which is kinematically forbidden within the conventional perturbative calculation and gives the dominant contribution in certain parameter regimes, e.g., light daughter particles. 
}

\begin{document} 
\maketitle

\section{Introduction}

Reheating is of great importance for successful inflation models to realize subsequent cosmological evolution such as Big Bang Nucleosynthesis.  In slow-roll inflation models, inflation ends when a (coherent) inflaton field falls down into its potential minimum.  The energy of the inflaton field is eventually converted into light elements, which completes reheating.  The very first stage of reheating is known to be dominated by the so-called preheating~\cite{Dolgov:1989us,Traschen:1990sw,Kofman:1994rk,Shtanov:1994ce,Yoshimura:1995gc,Kofman:1997yn} (see also, e.g. \cite{Bassett:2005xm,Allahverdi:2010xz,Amin:2014eta} for review), where relatively light matter species coupled directly to the inflaton field are explosively produced from the vacuum during the multiple oscillations of the inflaton field.  Lattice simulations are conveniently employed to analyze the preheating process~\cite{Khlebnikov:1996mc,Khlebnikov:1996wr,Prokopec:1996rr,Khlebnikov:1996zt}, as the amount of light particles produced during preheating is huge and hence the backreaction from the produced particles to the inflaton dynamics becomes important.

Instant preheating~\cite{Felder:1998vq} is proposed as a mechanism to accelerate the reheating process.   The idea is to produce heavy particles via decay of parent particles produced from the vacuum in the above-mentioned preheating mechanism.  The decay is thought to occur when the inflaton-dependent mass of the parent particle exceeds the daughter-particle mass because of the kinematic reason.  The decay then becomes the most efficient when the parent particle's mass reaches the (local) maximum.  Since the amplitude of the inflaton oscillation decays in time due to the expansion of the Universe, instant preheating may occur only within the first few oscillations of the inflaton.  As the amount of the produced particles after the first few oscillations is presumably small, the backreaction to the inflaton may not be large and the lattice simulation is not necessarily required for its description.  If the inflaton energy decays faster than that of non-relativistic matter, the energy of the heavy daughter particles eventually dominates the Universe, and thereby it completes reheating.  Instant preheating becomes particularly important, e.g., in quintessential models~\cite{Felder:1998vq,Felder:1999pv}.

From a general viewpoint, preheating is a quantum process driven by a coherent field.  Therefore, it should be formulated in the language of quantum-field theory (QFT) in a coherent field (or strong field).  In the literature, preheating is described as the dynamics of quantum fields coupled to a coherent field but without interactions between quantum fields.  More generally, we need a theory of interacting quantum fields in order to describe, e.g., instant preheating, which has not been fully addressed yet. In fact, the existing formulations of instant preheating were rather phenomenological.  They rely on cross sections derived with usual QFT perturbation theory {\it without} coherent fields, and the effects of coherent fields are included phenomenologically, e.g., via replacing a bare mass with a field-dependent mass by hand.  Such a phenomenological treatment clearly loses the information of multiple scatterings between coherent fields and particles during scattering processes.  Since the inflaton field is non-perturbatively strong in preheating, it is natural to expect that multiple-scattering effects become important.  The validity of the existing phenomenological formulations should thus be tested against a rigorous one.  As far as we know, such a rigorous QFT formulation of preheating has not been addressed in the literature.  We here tackle this problem.  We shall show how the multiple scatterings with the inflaton field change the particle production and clarify when the phenomenological formulations are valid.  

To provide a QFT formulation of instant preheating, we notice that there has been significant progress in the theory and phenomenology of QFT in a coherent field.  In particular, in the context of quantum electrodynamics (QED) and intense-laser physics \cite{DiPiazza:2011tq, Fedotov:2022ely}, a perturbation theory of particles dressed by coherent fields, known as the Furry-picture perturbation theory \cite{Furry:1951zz}, has been developed.  The Furry-picture perturbation theory has been established as a standard tool to analyze QED processes under coherent electromagnetic fields, including analogous processes to instant preheating (e.g., photon creation accompanied by the Schwinger pair production under a strong electric field \cite{Blaschke:2008wf,Blaschke:2011is,Tanji:2015ata,Otto:2016xpn,Aleksandrov:2019irn,Smolyansky:2019yma,Taya:2021dcz,Aleksandrov:2021ylw,Aleksandrov:2022rgg},  assistance to the particle production due to multiple-scattering by a strong electromagnetic field in the Schwinger pair production \cite{Schutzhold:2008pz, Dunne:2009gi, Taya:2018eng, Huang:2019uhf, Taya:2020bcd, Taya:2020pkm} and in the non-linear Breit-Wheeler process \cite{Nousch:2012xe, Titov:2012rd}).  This fact motivates us to think about the formulation of instant preheating by extending the Furry-picture perturbation theory in QED (i.e., a single charged fermion on top of an electromagnetic field in the flat spacetime) to instant preheating (i.e., two types of particles on top of an inflaton field in the expanding Universe).

The purpose of this paper is to formulate instant preheating based on QFT with the aid of the Furry-picture perturbation theory, 
which as far as we know has not been addressed in the literature.  Using the QFT formulation, we first examine the validity of the conventional phenomenological approach.  We show that the phenomenological approach is justified in a certain parameter regime (called ``kinematically-allowed regime''), provided that the inflaton field is sufficiently slow and weak.  We then point out that there exists another particle-producing parameter regime (called ``kinematically-forbidden regime'') which cannot be described by the conventional approach.  We shall show that the particle production in this parameter regime is driven by a novel non-perturbative mechanism assisted by the coherent inflaton field.  We discuss the behavior of the kinematically-forbidden process, which is far different from a perturbative ``decay process'' of parent particles without background coherent fields.  It turns out that whether the kinematically-allowed or the kinematically-forbidden process dominates depends on a parameter region under consideration, which indicates that our QFT formulation is imperative for a complete description of the history of the Universe.

This paper is organized as follows: We provide the QFT formulation of instant preheating based on the Furry-picture perturbation theory in Sec.~\ref{sec:2}.  We present a general analytical formula for the number density of produced particles, as well as a simpler one by assuming that the effective masses of dressed particles are sufficiently slow.  We also explain how the conventional perturbative approach is included in our QFT formula and point out the novel non-perturbative particle-production mechanism in the kinematically-forbidden regime.  In Sec.~\ref{sec3}, we consider a specific preheating model and discuss the behaviors of the number and energy density of daughter particles in more detail.  Finally, we give a summary and an outlook in Sec.~\ref{sec4}.

\vspace*{2mm}

\noindent{\it Notation}: Our spacetime $x^\mu = (t,{\bm x})$ is (1+3) dimensions.  We adopt the mostly plus metric convention, i.e., $g^{\mu\nu} = {\rm diag}(-1,+1,+1,+1)$ in the flat spacetime.  We adopt the natural unit $\hbar = c =1$.

\section{Furry-picture approach to instant preheating} \label{sec:2}

\begin{figure}[t]
\centering
\begin{tikzpicture}
\begin{feynhand}
	\vertex[particle]  (A) at (-2,0); 
	\vertex[particle]  (a0) at (-2.35,0); 
	\vertex[crossdot] (ac0) at (-2.8,0) {}; 
	\vertex[particle] (a1) at (-2.0,0.77); 
	\vertex[particle] (a2) at (-1.47,1.01); 
	\vertex[crossdot] (ac2) at (-1.73,+1.41) {}; 
	\vertex[particle] (a3) at (-0.93,1.2); 
	\vertex[crossdot] (ac3) at (-1.1,+1.63) {}; 
	\vertex[particle] (a4) at (-0.49,1.31); 
	\vertex[dot] (a5) at (-0.2,1.32); 
	\vertex[particle] (B) at (+2,0) {};
	\vertex[particle] (b1) at (-2.0,-0.77); 
	\vertex[particle] (b2) at (-1.49,-1.01); 
	\vertex[crossdot] (bc2) at (-1.68,-1.42) {}; 
	\vertex[particle] (b5) at (1.8,-1.42) {$\chi$}; 
	\vertex[particle] (b6) at (0.2,-1.37); 
	\vertex[crossdot] (bc6) at (0.15,-1.8) {}; 
	\vertex[particle] (b7) at (0.91,-1.41); 
	\vertex[crossdot] (bc7) at (0.93,-1.85) {}; 

	\vertex[particle] (d1) at (1.8,2.32) {$\psi$}; 
	\vertex[particle] (d2) at (1.8,0.32) {$\psi$}; 

	\vertex[particle] (l1)  at (-2.63,+0.59) {.};
	\vertex[particle] (l2)  at (-2.45,+0.9) {.};
	\vertex[particle] (l3)  at (-2.20,+1.13) {.};
	\vertex[particle] (l4)  at (-2.63,-0.59) {.};
	\vertex[particle] (l5)  at (-2.45,-0.95) {.};
	\vertex[particle] (l6)  at (-2.15,-1.16) {.};
	\vertex[particle] (l7)  at (-1.18,-1.6) {.};
	\vertex[particle] (l8)  at (-0.79,-1.705) {.};
	\vertex[particle] (l9)  at (-0.40,-1.73) {.};
	
	\propag[plain] (a5) to [in=90, out=180, looseness=0.8] (a0);
	\propag[plain] (a0) to [in=180, out=-90, looseness=0.8] (b5);
	\propag[photon] (ac0) to (a0);
	\propag[photon] (ac2) to (a2);
	\propag[photon] (ac3) to (a3);
	\propag[photon] (bc2) to (b2);
	\propag[photon] (bc6) to (b6);
	\propag[photon] (bc7) to (b7);
	\propag[scalar] (a5) to (d1);
	\propag[scalar] (a5) to (d2);
\end{feynhand}
\end{tikzpicture}
\caption{A diagrammatic picture of instant preheating.  A pair of $\chi$ particles (solid line) are produced from the vacuum, one of which subsequently decays into a pair of $\psi$ particles (dashed lines).  The parent $\chi$ particles couple to the coherent $\phi$ field (blobs) in a non-perturbative manner, while the daughter $\psi$ particles do not.  }\label{fig1}
\end{figure}
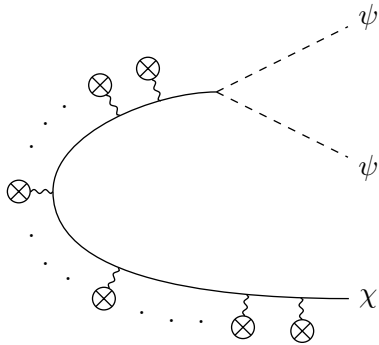 

We provide a quantum field-theoretic formulation of instant preheating in the early Universe based on the Furry-picture perturbation theory~\cite{Furry:1951zz}.  

We derive a general formula for the number density of particles interacting with each other under the presence of a coherent background inflaton field (see Fig.~\ref{fig1}).  Assuming that the inflaton field is sufficiently slow in time, we explicitly evaluate the general formula to obtain a compact expression for the number of daughter particles.  The compact formula describes not only the conventional perturbative particle production, which is obtained by the standard perturbation theory with {\it un}-dressed particles, but also the non-perturbative one in the kinetically-forbidden regime that the conventional approach cannot access.

\subsection{Model}

For concreteness, we consider a particular cosmological model of preheating (although the Furry-picture perturbation theory is a general framework that can be applied to a variety of models that include, e.g., those with fermions and gauge bosons).  Our model consists of two real scalar fields $\chi$ and $\psi$, interacting with each other via a cubic interaction, under the expanding Universe and a coherent background real scalar field $\phi$, representing an inflaton.  The action $S$ is
\begin{align}
     S = -\int {\rm d}^4x\sqrt{-g}\left[\frac12(\partial \chi)^2+\frac12m^2_\chi\chi^2+\frac12\zeta\chi^2\phi^2+\frac12(\partial\psi)^2+\frac12m_\psi^2\psi^2+\frac12\lambda\chi\psi^2\right] \;, \label{eq:2.1}
\end{align}
where $m_{\chi,\psi}, \zeta$, and $\lambda$ are constants, representing ``bare'' masses of $\chi,\psi$ fields, couplings between $\chi$ and $\phi$ and between $\chi$ and $\psi$, respectively, and $g$ is the determinant of the Friedmann-Robertson-Walker (FRW) metric ${\rm d}s^2=g_{\mu\nu} {\rm d}x^\mu {\rm d}x^\nu = -{\rm d}t^2+a^2(t){\rm d}{\bm x}^2$.  We have implicitly assumed that the backreaction by the dynamical $\chi,\psi$ fields onto $\phi$ and the metric $g_{\mu\nu}$ (or the scale factor $a$) are negligible so that they can be treated as non-dynamical backgrounds.  Such a treatment may be justified in the early stage of preheating, if the work done by the backgrounds $\phi$ and $a(t)$ to produce $\chi,\psi$ particles are sufficiently small compared to the energies that the backgrounds originally possess.

It is convenient to redefine the dynamical fields as $a^{-3/2} \chi \to \chi$ and $a^{-3/2}\psi \to \psi$ to express the action (\ref{eq:2.1}) as
\begin{align}
    S \to \frac12 \int {\rm d}t{\rm d}^3{\bm x}\left[\dot\chi^2-\frac{1}{a^2}({\bm \partial}\chi)^2-M_\chi^2\chi^2+\dot\psi^2-\frac{1}{a^2}({\bm \partial}\psi)^2-M^2_{\psi}\psi^2-\lambda a^{-\frac32}\chi\psi^2\right] \;,
\end{align}
where $\dot{\bullet} := \partial_t \bullet$ and 
\begin{align}
\begin{split}
    M_\chi^2(t) &:= m_\chi^2+\zeta\phi^2(t)-\frac94 H^2(t)-\frac32 \dot{H}(t) \;, \\
    M_\psi^2(t) &:= m_\psi^2-\frac94 H^2(t)-\frac32 \dot{H}(t) \;,\label{mass formula}
\end{split}
\end{align}
are ``dressed'' masses of $\chi, \psi$ fields and $H=\dot{a}/a$ is the Hubble parameter.  The background field $\phi$ is now absorbed into the dressed mass $M_\chi$, while it does not modify $M_\psi$ due to the absence of the coupling between $\psi$ and $\phi$.  The role of $\phi$ is thus to give time dependence to the mass of $\chi$ particles.  This time dependence is the essence of the vacuum pair production of $\chi$ particles and also affects the $\psi$ production through the coupling with $\chi$, as we explain in detail below.

Note that we assume later to get a simple expression that the bare masses $m_\chi$ and $m_\psi$ are larger than the Hubble scale, so that one may safely neglect the Hubble-induced-mass contributions to the particle-production processes [see Eq.~(\ref{BCcond})].  (Nonetheless, our general formulation until Eq.~(\ref{BCcond}) is applicable to arbitrary dressed masses that include the Hubble-induced contributions.  )  This assumption is equivalent to treating $H \sim 0$, or $a \sim {\rm const}.$, during a scattering process, and then the dressed masses \eqref{mass formula} can be approximated as $M_\chi^2 \sim m_\chi^2 + \zeta \phi^2$ and $M_\psi^2 \sim m_\psi^2$.

\subsection{The Furry-picture perturbation theory}\label{Furry}

We formulate instant preheating based on the Furry-picture perturbation theory and derive a general formula for the numbers of $\chi$ and $\psi$ particles produced from an initial vacuum.  Note that although we focus on the number of particles throughout this paper, it is straightforward to extend our formalism to compute other physical quantities such as energy-momentum tensor.

The idea of the Furry-picture perturbation theory is to treat interactions between dynamical $\chi, \psi$ fields and the backgrounds $\phi$ and $a(t)$ non-perturbatively while those among dynamical fields perturbatively.  To achieve this, it is convenient to split the Hamiltonian ${\mathcal H}$ for the action $S$ into two parts: 
\begin{align}
    {\mathcal H}(t) = {\mathcal H}^{(0)}(t) + {\mathcal H}^{\rm int}(t) \;,
\end{align}
where
\begin{align}
\begin{split}
    &{\mathcal H}^{(0)} := \int {\rm d}^3{\bm x}\left[\dot\chi^2+\frac{1}{a^2}({\bm \partial}\chi)^2+M_\chi^2\chi^2+\dot\psi^2+\frac{1}{a^2}({\bm \partial}\psi)^2+M^2_{\psi}\psi^2\right] \;, \\
    &{\mathcal H}^{\rm int} := \lambda \int {\rm d}^3{\bm x}\, a^{-\frac32}\chi\psi^2 \;,
\end{split}
\end{align}
describing interactions between the dynamical fields and the backgrounds and those among the dynamical fields, respectively.

We evaluate the dynamical $\chi,\psi$ fields non-perturbatively in terms of ${\mathcal H}^{(0)}$.  The solutions $\chi^{(0)}$ and $\psi^{(0)}$ are not simple free fields but are dressed by the backgrounds $\phi$ and $a(t)$.  In other words, the mode functions in the momentum space $\chi^{(0)}_{\bm p}$ and $\psi^{(0)}_{\bm k}$ such that (note that we consistently use ${\bm p}$ and ${\bm k}$ to label momenta of $\chi$ and $\psi$ particles, respectively)  
\begin{align}
\begin{split}
    \chi^{(0)}(t,{\bm x}) 
        = \int\frac{{\rm d}^3{\bm p}}{(2\pi)^{\frac32}}{\rm e}^{{\rm i}{\bm p}\cdot{\bm x}}\left[\chi^{(0)}_{\bm p}(t)\hat{a}_{\bm p}^{\rm in}+\chi^{(0)*}_{-{\bm p}}(t)\hat{a}^{{\rm in}\dagger}_{-\bm p}\right] \;, \\
    \psi^{(0)}(t,{\bm x}) 
        = \int\frac{{\rm d}^3{\bm k}}{(2\pi)^{\frac32}}{\rm e}^{{\rm i}{\bm k}\cdot{\bm x}}\left[\psi^{(0)}_{\bm k}(t)\hat{b}^{\rm in}_{\bm k}+\psi^{(0)*}_{-{\bm k}}(t)\hat{b}^{{\rm in}\dagger}_{-\bm k}\right] \;,
\end{split}
\end{align}
are not usual plane waves but are dressed so that they satisfy the Klein-Gordon equations with time-dependent mass terms,
\begin{align}
    0 = \ddot{\chi}^{(0)}_{\bm p} + \omega_{\bm p}^2 \chi^{(0)}_{\bm p} \;, \quad
    0 = \ddot{\psi}^{(0)}_{\bm k} + \Omega_{\bm k}^2 \psi^{(0)}_{\bm k} \;, \label{eq:2.7}
\end{align}
with
\begin{align}
    \omega_{\bm p}^2 := M_\chi^2+{\bm p}^2/a^2  \;, \quad
    \Omega_{\bm k}^2 := M_\psi^2+{\bm k}^2/a^2 \;.
\end{align}
To uniquely fix the mode functions $\chi^{(0)}_{\bm p}$ and $\psi^{(0)}_{\bm k}$, one needs to specify the boundary and normalization conditions.  To do so, we formally assume that the backgrounds are turned off at an initial time $t=t_{\rm in}$, so that one can naturally identify positive frequency mode at the in-state $t=t_{\rm in}$ with the usual plane waves.  Thus, we require\footnote{Strictly speaking, the right-hand sides of Eq.~(\ref{eq:2.9}) are the lowest-order Wentzel-Kramers-Brillouin (WKB) mode functions and are not the usual plane waves.  The lowest-order WKB mode functions and the usual plane waves exactly agree with each other in the regime where the time dependence of the backgrounds is turned off.  In this paper, therefore, we do not carefully distinguish the lowest-order WKB mode function and the plane waves in the asymptotically free in- and out-states.  }
\begin{align}
\begin{split}
    \chi^{(0)}_{\bm p}(t) &\xrightarrow{t\to t_{\rm in}} \frac{1}{\sqrt{2\omega_{\bm p}(t)}}{\rm e}^{-{\rm i}\int^t {\rm d}t'\,\omega_{\bm p}(t')}  \; , \label{eq:2.9} \\
    \psi^{(0)}_{\bm k}(t) &\xrightarrow{t\to t_{\rm in}} \frac{1}{\sqrt{2\Omega_{\bm k}(t)}}{\rm e}^{-{\rm i}\int^t {\rm d}t'\,\Omega_{\bm k}(t')} \; .
\end{split}
\end{align}
Note that the lower end of the integral in the exponent can be anything (as long as it is real), as a different lower end just changes the unimportant overall phase of the mode function.  We here implicitly normalized the mode functions as
\begin{align}
    1 = -{\rm i}\chi^{(0)}_{\bm p} \overleftrightarrow{\partial_t} \chi^{(0)*}_{\bm p} = -{\rm i} \psi^{(0)}_{\bm k} \overleftrightarrow{\partial_t} \psi^{(0)*}_{\bm k} \;,\label{norm}
\end{align}
with $\overleftrightarrow{\partial_t} := \overrightarrow{\partial_t} - \overleftarrow{\partial_t}$.  The normalization condition (\ref{norm}) in turn normalizes the canonical commutation relations for the annihilation operators $\hat{a}^{\rm in}_{\bm p},\hat{b}^{\rm in}_{\bm k}$ as
\begin{align}
    \delta({\bm p}-{\bm p}') = [\hat{a}^{\rm in}_{\bm p},\hat{a}^{{\rm in}\dagger}_{{\bm p}'}] \;, \quad
    \delta({\bm k}-{\bm k}') = [\hat{b}^{\rm in}_{\bm k},\hat{b}^{{\rm in}\dagger}_{{\bm k}'}] \;. 
\end{align}
Requiring the boundary condition (\ref{eq:2.9}) is equivalent to requiring that the annihilation operators $\hat{a}^{\rm in}_{\bm k} $ and $\hat{b}^{\rm in}_{\bm k}$ coincide at the in-state $t=t_{\rm in}$ with the usual asymptotic annihilation operators: 
\begin{align}
\begin{split}
    \hat{a}^{\rm in}_{\bm p} 
        &= +{\rm i} \int {\rm d}^3{\bm x} \left( \frac{{\rm e}^{{\rm i}{\bm p}\cdot{\bm x}}}{(2\pi)^{\frac{3}{2}}} \chi^{(0)}_{\bm p} \right)^* \overleftrightarrow{\partial_t} \chi^{(0)}
        = \lim_{t\to t_{\rm in}} (+{\rm i}) \int {\rm d}^3{\bm x} \left( \frac{{\rm e}^{{\rm i}{\bm p}\cdot{\bm x}}}{(2\pi)^{\frac{3}{2}}} \frac{{\rm e}^{-{\rm i}\int^t {\rm d}t'\,\omega_{\bm p}(t')}}{\sqrt{2\omega_{\bm p}(t)}} \right)^* \overleftrightarrow{\partial_t} \chi^{(0)}   \; , \label{eq:2.11} \\
    \hat{b}^{\rm in}_{\bm k} 
        &= +{\rm i} \int {\rm d}^3{\bm x} \left( \frac{{\rm e}^{{\rm i}{\bm k}\cdot{\bm x}}}{(2\pi)^{\frac{3}{2}}} \psi^{(0)}_{\bm k} \right)^* \overleftrightarrow{\partial_t} \psi^{(0)}
        = \lim_{t\to t_{\rm in}} (+{\rm i}) \int {\rm d}^3{\bm x} \left( \frac{{\rm e}^{{\rm i}{\bm k}\cdot{\bm x}}}{(2\pi)^{\frac{3}{2}}} \frac{{\rm e}^{-{\rm i}\int^t {\rm d}t'\,\Omega_{\bm k}(t')}}{\sqrt{2\Omega_{\bm k}(t)}} \right)^* \overleftrightarrow{\partial_t} \psi^{(0)}   \; .
\end{split}
\end{align}
The corresponding vacuum state $\ket{0}_{\rm in}$, 
\begin{align}
    0 = \hat{a}^{\rm in}_{\bm p} \ket{0}_{\rm in} = \hat{b}^{\rm in}_{\bm k} \ket{0}_{\rm in}\ {\rm with}\ 1={}_{\rm in}\langle0|0\rangle_{\rm in} \; ,
\end{align}
is thus identical to the usual perturbative vacuum at the in-state $t=t_{\rm in}$.  We remark that the second equalities in Eq.~(\ref{eq:2.11}) hold only for $t \to t_{\rm in}$ in the presence of backgrounds (and/or interactions), since the dressed mode functions $\chi^{(0)}_{\bm p}$ and $\psi^{(0)}_{\bm k}$ coincide with the plane waves only at $t=t_{\rm in}$ and positive and negative frequency modes shall be mixed up with each other during the time evolution.  This means that asymptotic annihilation operators at the out-state $t\to+\infty$ do not coincide with those at the in-state $t=t_{\rm in}$.  As we see shortly below, this is the essence of particle production from the vacuum.  

We turn to include effects of the interactions among the dynamical fields due to ${\mathcal H}^{\rm int}$.  We formally assume that ${\mathcal H}^{\rm int}$ is turned on adiabatically at the initial time $t=t_{\rm in}$ and since then perturbatively evolve the field operators $\chi^{(0)} \to \chi$ and $\psi^{(0)} \to \psi$.\footnote{A special case of the Furry-picture perturbation theory is common in the calculation of cosmological correlators. There, the mode functions are non-perturbatively dressed by the de Sitter background, while interactions among dynamical fields are treated perturbatively~\cite{Weinberg:2005vy}.}  It is convenient to move to an interaction picture in which the time-translation unitary operator $U$ is constructed perturbatively with the interaction Hamiltonian ${\mathcal H}^{\rm int} = {\mathcal O}(\lambda)$ as (see e.g. Ref.~\cite{Weinberg:2005vy}),
\begin{align}
    U(t,t_{\rm in}) 
    = {\rm T}\exp\left(-{\rm i}\int_{t_0}^t {\rm d}t'\, {\mathcal H}^{\rm int}(t') \right) 
    = 1 -{\rm i}\int_{t_{\rm in}}^t {\rm d}t'\, {\mathcal H}^{\rm int}(t') + {\mathcal O}(\lambda^2) \;. \label{eq:2.13}
\end{align}
The time evolution of the dynamical $\chi, \psi$ fields can be expressed as
\begin{align}
    \begin{pmatrix} \chi(t) \\ \psi(t) \end{pmatrix} &= [U(t,t_{\rm in})]^\dagger \begin{pmatrix} \chi^{(0)}(t) \\ \psi^{(0)}(t) \end{pmatrix} U(t,t_{\rm in}) \; .
\end{align}
Inserting Eq.~(\ref{eq:2.13}), we get
\begin{align}
\begin{split}
    \chi(t,\bm x) 
        &= \chi^{(0)}(t,{\bm x})-{\rm i}\frac{\lambda}{2}\int_{t_{\rm in}}^t {\rm d}t'{\rm d}^3{\bm x}'a^{-\frac{3}{2}}(t')[\chi^{(0)}(t,{\bm x}),\chi^{(0)}(t',{\bm x}')]\psi^{(0)2}(t',{\bm x}') + \mathcal{O}(\lambda^2)\;, \label{eq:2.14}\\
    \psi(t,{\bm x})
        &= \psi^{(0)}(t,{\bm x})-{\rm i}\lambda\int_{t_{\rm in}}^t{\rm d}t'{\rm d}^3{\bm x}'a^{-\frac{3}{2}}(t')[\psi^{(0)}(t,{\bm x}),\psi^{(0)}(t',{\bm x}')]\chi^{(0)}(t',{\bm x}')\psi^{(0)}(t',{\bm x}') + \mathcal{O}(\lambda^2)\;. 
\end{split}
\end{align}
Note that the field renormalization factor is higher order in the coupling $\lambda$ as $Z_{\chi,\psi}=1+\mathcal{O}(\lambda^2)$ and hence can be neglected safely at the present accuracy.

We wish to compute the numbers of $\chi, \psi$ particles at the out-state $t\to+\infty$ produced from the initial vacuum $\ket{0}_{\rm in}$.  The phase-space densities at each momentum mode, which we write $n^\chi_{\bm p}$ and $n^\psi_{\bm k}$, are defined as 
\begin{align}
     n^\chi_{\bm p} 
        :=  \frac{ {}_{\rm in}\langle0|\hat{a}^{{\rm out}\dagger}_{\bm p}\hat{a}^{\rm out}_{\bm p}|0\rangle_{\rm in} }{\delta^3({\bm p}={\bm 0})}\; ,\quad
     n^\psi_{\bm k} 
        := \frac{ {}_{\rm in}\langle0|\hat{b}^{{\rm out}\dagger}_{\bm k}\hat{b}^{\rm out}_{\bm k}|0\rangle_{\rm in} }{\delta^3({\bm k}={\bm 0})} \; , \label{eq:2.20}
\end{align}
where $\delta^3({\bm p}={\bm 0}) = \delta^3({\bm k}={\bm 0}) =: V/(2\pi)^3$ with $V$ the whole spatial volume.  The annihilation operators at the out-state $\hat{a}^{\rm out}_{\bm p}$ and $\hat{b}^{\rm out}_{\bm k}$ are defined in a similar manner as the in-state operators $\hat{a}^{\rm in}_{\bm p}$ and $\hat{b}^{\rm in}_{\bm k}$ (\ref{eq:2.11}).  Namely, we formally assume that the backgrounds and the interactions are switched off adiabatically at the out-state $t\to\infty$.  We can then naturally identify positive-frequency modes at the out-state with the usual plane waves.  Thus, we introduce the annihilation operators at the out-state $\hat{a}^{\rm out}_{\bm p}$ and $\hat{b}^{\rm out}_{\bm k}$ as
\begin{align}
\begin{split}
   \hat{a}^{\rm out}_{\bm p} 
        &:= \lim_{t\to +\infty} (+{\rm i}) \int {\rm d}^3{\bm x} \left( \frac{{\rm e}^{{\rm i}{\bm p}\cdot{\bm x}}}{(2\pi)^{\frac{3}{2}}} \frac{{\rm e}^{-{\rm i}\int^t {\rm d}t'\,\omega_{\bm p}(t')}}{\sqrt{2\omega_{\bm p}(t)}} \right)^* \overleftrightarrow{\partial_t} \chi   \; , \label{eq:2.17}\\
    \hat{b}^{\rm out}_{\bm k} 
        &:= \lim_{t\to +\infty} (+{\rm i}) \int {\rm d}^3{\bm x} \left( \frac{{\rm e}^{{\rm i}{\bm k}\cdot{\bm x}}}{(2\pi)^{\frac{3}{2}}} \frac{{\rm e}^{-{\rm i}\int^t {\rm d}t'\,\Omega_{\bm k}(t')}}{\sqrt{2\Omega_{\bm k}(t)}} \right)^* \overleftrightarrow{\partial_t} \psi   \; .
\end{split}
\end{align}
The corresponding out-vacuum state $\ket{0}_{\rm out}$ is defined as a state such that
\begin{align}
    0 = \hat{a}^{\rm out}_{\bm p} \ket{0}_{\rm out} = \hat{b}^{\rm out}_{\bm k} \ket{0}_{\rm out} \ {\rm with}\ 1={}_{\rm out}\langle0|0\rangle_{\rm out}\;.
\end{align}
We emphasize that the out-vacuum $\ket{0}_{\rm out}$ does not coincide with the in-vacuum $1\neq{}_{\rm in}\langle0|0\rangle_{\rm out}$ due to the interactions in ${\mathcal H}_0$ and ${\mathcal H}_{\rm int}$.  The inequivalence $\ket{0}_{\rm out} \neq \ket{0}_{\rm in}$ is a consequence of $\hat{a}^{\rm out}_{\bm p} \neq \hat{a}^{\rm in}_{\bm p}$ and $\hat{b}^{\rm out}_{\bm k} \neq \hat{b}^{\rm in}_{\bm k}$.  The relationship between the in- and out-state annihilation operators can be obtained, by plugging Eq.~(\ref{eq:2.14}) into Eq.~(\ref{eq:2.17}), as
\begin{align}
\begin{split}
    \hat{a}^{\rm out}_{\bm p}  
        &= \alpha_{{\bm p}}\hat{a}^{\rm in}_{\bm p}+\beta_{{\bm p}}^*\hat{a}_{-\bm p}^{{\rm in}\dagger} \\
        &\quad - {\rm i} \frac{\lambda}{2}\int_{t_{\rm in}}^\infty {\rm d}t'{\rm d}^3{\bm x}'a^{-\frac32}(t') {\rm e}^{-{\rm i}{\bm p}\cdot{\bm x}'} \left( \alpha_{\bm p}\chi^{(0)*}_{{\bm p}}(t') - \beta_{-{\bm p}}^*\chi^{(0)}_{-{\bm p}}(t') \right) \left(\psi^{(0)}(t',{\bm x}')\right)^2 + \mathcal{O}(\lambda^2) \;,  \label{eq:2.19} \\
    \hat{b}^{\rm out}_{{\bm k}} 
        &= \gamma_{{\bm k}}\hat{b}^{\rm in}_{\bm k}+\delta_{{\bm k}}^*\hat{b}_{-{\bm k}}^{{\rm in}\dagger} \\
        &\quad -{\rm i} \lambda\int_{t_{\rm in}}^\infty {\rm d}t'{\rm d}^3{\bm x}'a^{-\frac32}(t') {\rm e}^{-{\rm i}{\bm k}\cdot{\bm x}'}\left( \gamma_{{\bm k}} \psi^{(0)*}_{{\bm k}}(t') - \delta_{-{\bm k}}^* \psi^{(0)}_{-{\bm k}}(t') \right) \psi^{(0)}(t',{\bm x}') \chi^{(0)}(t',{\bm x}') + \mathcal{O}(\lambda^2) \;,
\end{split}
\end{align}
where 
\begin{align}
\begin{split}
    \begin{pmatrix} \alpha_{\bm p} \\ \beta_{\bm p} \end{pmatrix}
        &= \lim_{t\to+\infty} (+{\rm i}) \begin{pmatrix} \frac{{\rm e}^{+{\rm i}\int^t {\rm d}t'\,\omega_{\bm p}(t')}}{\sqrt{2\omega_{\bm p}(t)}} \\ -\frac{{\rm e}^{-{\rm i}\int^t {\rm d}t'\,\omega_{\bm p}(t')}}{\sqrt{2\omega_{\bm p}(t)}} \end{pmatrix} \overleftrightarrow{\partial_t} \chi^{(0)}_{\bm p} \; , \label{eq:2.20} \\
    \begin{pmatrix} \gamma_{\bm k} \\ \delta_{\bm k} \end{pmatrix}
        &= \lim_{t\to+\infty} (+{\rm i}) \begin{pmatrix} \frac{{\rm e}^{+{\rm i}\int^t {\rm d}t'\,\Omega_{\bm k}(t')}}{\sqrt{2\Omega_{\bm k}(t)}} \\ -\frac{{\rm e}^{-{\rm i}\int^t {\rm d}t'\,\Omega_{\bm k}(t')}}{\sqrt{2\Omega_{\bm k}(t)}} \end{pmatrix} \overleftrightarrow{\partial_t} \psi^{(0)}_{\bm k} \; ,
\end{split}
\end{align}
are called Bogoliubov coefficients.  Note that the Bogoliubov coefficients satisfy $1 = |\alpha_{\bm p}|^2-|\beta_{\bm p}|^2 = |\gamma_{\bm k}|^2-|\delta_{\bm k}|^2$, which follows directly from the normalization conditions of the mode functions \eqref{norm}.   Plugging Eq.~(\ref{eq:2.19}) into Eq.~(\ref{eq:2.20}), we find for ${\bm 0} \neq {\bm p},{\bm k}$ that  
\begin{align}
\begin{split}
    n^\chi_{\bm p} 
        &= [\delta^3({\bm p}={\bm 0})]^{-1} {}_{\rm in}\langle0|\Bigl|\beta_{{\bm p}}^*\hat{a}_{-{\bm p}}^{{\rm in}\dagger} \\
        &\quad - {\rm i} \frac{\lambda}{2} \int_{t_{\rm in}}^\infty {\rm d}t' \int{\rm d}^3{\bm k} \, a^{-\frac{3}{2}}(t') \left( \alpha_{{\bm p}} \chi^{(0)*}_{{\bm p}}(t') - \beta_{-{\bm p}}^* \chi^{(0)}_{-{\bm p}}(t') \right)  \psi^{(0)*}_{-{\bm k}}(t') \psi^{(0)*}_{\bm k-\bm p}(t') \hat{b}^{{\rm in}\dagger}_{-{\bm k}} \hat{b}^{{\rm in}\dagger}_{{\bm k}-{\bm p}} + \mathcal{O}(\lambda^2)  \Bigr|^2 |0\rangle_{\rm in}  \; ,  
        \\
     n^\psi_{\bm k} 
        &= [\delta^3({\bm k}={\bm 0})]^{-1} {}_{\rm in}\langle0|\Bigl|\delta_{{\bm k}}^*\hat{b}_{-{\bm k}}^{{\rm in}\dagger} \\
        &\quad -{\rm i} \lambda\int_{t_{\rm in}}^\infty {\rm d}t' \int{\rm d}^3{\bm p} \, a^{-\frac{3}{2}}(t') \left( \gamma_{{\bm k}} \psi^{(0)*}_{{\bm k}}(t') - \delta_{-{\bm k}}^* \psi^{(0)}_{-{\bm k}}(t') \right) \chi^{(0)*}_{\bm p}(t) \psi^{(0)*}_{-{\bm k}-{\bm p}}(t')\hat{a}^{{\rm in}\dagger}_{{\bm p}} \hat{b}^{{\rm in}\dagger}_{-\bm k-\bm p} + \mathcal{O}(\lambda^2) \Bigr|^2 |0\rangle_{\rm in} \; . \label{eq:2.21}
\end{split}
\end{align}
In the limit of $\lambda \to 0$, only the off-diagonal Bogoliubov coefficients $\beta_{\bm p}$ and $\delta_{\bm k}$ [i.e., the first terms in Eq.~(\ref{eq:2.21})] can contribute to the particle production.  This physically means that the particle production is driven by the background $\phi$ as well as $a(t)$.  For $\lambda \neq 0$, particles can also be produced via scattering/radiation processes among the dynamical $\chi,\psi$ fields, which are described by the contributions from the second terms in Eq.~ (\ref{eq:2.21}).

Below, we focus on the production of $\psi$ particles, which is of particular interest in instant preheating.  We use the assumption that the Hubble mass terms in the dressed masses \eqref{mass formula} are negligible, so that only the inflaton $\phi$ field can affect the dressed mass.  This means that the Bogoliubov coefficients can be non-trivial only for the $\chi$ field, as only the $\chi$ field couples to the $\phi$ field, i.e., 
\begin{align}
    \alpha_{\bm p}\neq1\;,\quad \beta_{\bm p}\neq0\;,\quad \gamma_{\bm k}=\;1,\quad \delta_{\bm k}=0 \; . \label{BCcond}
\end{align}
Plugging this expression into $n^\psi_{\bm k}$ (\ref{eq:2.20}) leads to
\begin{align}
    n^\psi_{\bm k}
        & = \lambda^2\int\frac{{\rm d}^3{\bm p}}{(2\pi)^3}\left|\int^{\infty}_{t_{\rm in}}{\rm d}t'a^{-\frac32}(t')\chi^{(0)}_{\bm p}(t')\psi^{(0)}_{-{\bm k}-{\bm p}}(t')\psi^{(0)}_{{\bm k}}(t') + {\mathcal O}(\lambda) \right|^2 \; . \label{npk}
\end{align}

Remark that Eq.~(\ref{npk}) is rather formal at this stage: To calculate the actual number density (\ref{npk}), one needs to (i) know the dressed mode functions $\chi^{(0)}_{\bm p}$ and $\psi^{(0)}_{\bm k}$ satisfying the mode equation (\ref{eq:2.7}); and (ii) carry out the time integral with $\chi^{(0)}_{\bm p}$ and $\psi^{(0)}_{\bm k}$.  In practice, the steps (i) and (ii) are not necessarily easy.  In fact, for (i), analytical solutions to the mode equation (\ref{eq:2.7}) are not available for general backgrounds.  Thus, one needs to rely on numerical methods to solve Eq.~(\ref{eq:2.7}) mode-by-mode, which is not necessarily cheap from a numerical point of view.  For (ii), the integrand is a highly oscillating function in time, which is difficult to be integrated even numerically.  To overcome those problems (i) and (ii), we need some approximation.  As we describe in Sec.~\ref{sec2.3}, we shall further assume in this paper that the background $\phi$ is also slow in time (but gives a larger contribution than the Hubble terms to the effective mass that leads to non-trivial effects such as the vacuum $\chi$ particle production).  In such a limit, the mode equation (\ref{eq:2.7}) and also the resulting integral can be evaluated exactly, and thereby the problems (i) and (ii) can be resolved.

\subsection{Non-perturbative evaluation in slow-field limit}
\label{sec2.3}

To explicitly calculate the number of $\psi$ particles~\eqref{npk}, we make two further assumptions: (i) The background $\phi$ is varying very slowly, i.e., the typical frequency of $\phi$ is much smaller than the other scales of the problem such as the masses of the particles.  (ii) The time when the production of parent $\chi$ particles occurs dominantly, which we write $t_{\rm prod}$, is sufficiently later than the initial time $t_{\rm in} \ll t_{\rm prod}$, so that one can safely take $t_{\rm in} \to -\infty$. 

From the assumption (i), it is legitimate to Taylor expand the backgrounds around the production time $t=t_{\rm prod}$ (cf. in reheating, it is an expansion of the inflaton field $\phi$ around its first minimum, where the energy cost to create a pair is minimized\footnote{\label{foot3}Mathematically, vacuum pair production occurs when the Stokes line in the complex time plane crosses the real axis (see e.g. Refs.~\cite{Dumlu:2010ua,Dumlu:2010vv,Dumlu:2011cc,Dumlu:2011rr,Enomoto:2020xlf,Taya:2020dco,Hashiba:2021npn,Yamada:2021kqw}), and the crossing time is identified to be $t_{\rm prod}$.  (Note that the Stokes-line crossing does not necessarily occur at when the adiabaticity condition is maximally violated.  The adiabaticity condition determines the probability of a production event at a {\it given} production time and is in principle an independent condition to determine the crossing time.)  Stokes lines are emanating from the so-called turning points $t_{\rm tp}$ such that $\omega_{\bm p}(t_{\rm tp}) = 0$.  It is natural to expect that the turning points are located close to where $\phi$, and accordingly $\omega_{\bm p}$, is minimized (i.e., the closest to zero) on the real-time axis.  Then, the associated Stokes-line crossing should also occur at around the minima of $\phi$.  Thus, one may estimate $t_{\rm prod}$ as the time when $\phi$ is minimized.  Although this is a rough estimate, one can actually analyze the Stokes-line structure for the model (\ref{phsetup}) to be discussed in Sec.~\ref{sec3.2} and show that the Stokes-line crossings precisely occur at the minima of $\phi$.  Nonetheless, we note in general that for more complex models the production time could deviate from the minima of $\phi$ and the precise determination of which requires a careful analysis of the Stokes-line structure for each model.  This is beyond the scope of the present work.  }
). Keeping up to ${\mathcal O}(|t-t_{\rm prod}|^2)$ terms of $\phi^2$ [while $a$ is treated as a constant, as we have assumed below Eq.~(\ref{mass formula})]
\begin{align}
    a(t)=a(t_{\rm prod})\;, \quad \phi^2(t) = 
    \zeta^{-1}(A(t-t_{\rm prod}))^2 \;, \label{wsetup}
\end{align}
where $A$ is a real constant and we have assumed $\phi(t_{\rm prod})=0$ for simplicity\footnote{This assumption means that the vacuum expectation value of $\phi$ is zero. If this is not the case, we would find a cubic coupling $\sim \phi\chi^2$ in the effective Lagrangian.}.  The corresponding frequencies $\omega_{\bm p}$ and $\Omega_{\bm k}$ are, respectively, given by
\begin{align}
    \omega_{\bm p}^2(t) = \mu_{\bm p}^2 + (A(t-t_{\rm prod}))^2\;,\quad 
    \Omega_{\bm k}^2 = \frac{{\bm k}^2}{a^2(t_{\rm prod})} + m_\psi^2\;, \label{womega}
\end{align}
where
\begin{align}
    \mu_{\bm p}^2 
        := (M^{\rm min}_\chi)^2 + \frac{{\bm p}^2}{a^2(t_{\rm prod})} 
        := m_\chi^2 
        + \frac{{\bm p}^2}{a^2(t_{\rm prod})}  \; . 
\end{align}
The role of the scale factor $a(t_{\rm prod})$ is now just to rescale the momenta.  The dressed mass of $\chi$ particles $M_{\chi}(t) = \sqrt{(M^{\rm min}_\chi)^2 + (A(t-t_{\rm prod}))^2}$ takes its minimum $M^{\rm min}_\chi = m_\chi$ at $t=t_{\rm prod}$ and then can grow infinitely large as time goes within our approximation (\ref{wsetup}).  The infinitely large mass may cause unphysical behaviors, e.g., divergence in the total number $\int {\rm d}^3{\bm k}\,n^\psi_{\bm k}$.  Such unphysical behaviors are artifacts due to the Taylor expansion.  To remove the artifacts, one needs to introduce a proper cutoff (see Sec.~\ref{sec3}).  

Under the assumptions (i) and (ii), one can exactly solve the mode equation (\ref{eq:2.7}) to obtain the dressed mode functions $\chi^{(0)}_{\bm p}$ and $\psi^{(0)}_{\bm k}$.  Since $\Omega_{\bm k}$ (\ref{womega}) is time-independent owing to the assumption (i), the mode equation (\ref{eq:2.7}) for $\psi^{(0)}_{\bm k}$ can be solved trivially to give a plane wave: 
\begin{align}
    \psi^{(0)}_{\bm k} = \frac{1}{\sqrt{2\Omega_{\bm k}}} {\rm e}^{-{\rm i}\Omega_{\bm k}t} \; .
\end{align}
The mode equation (\ref{eq:2.7}) for $\chi^{(0)}_{\bm p}$ takes a rather non-trivial form due to the time dependence of $\omega_{\bm p}$ (\ref{womega}), 
\begin{align}
    \ddot{\chi}^{(0)}_{\bm p}(t) + \left( \mu_{\bm p}^2 + (A(t-t_{\rm prod}))^2 \right)\chi^{(0)}_{\bm p}(t) = 0 \;,
\end{align}
but is an analytically solvable equation (Weber equation).  Requiring the in-state boundary condition (\ref{eq:2.9}) and using the assumption (ii) (i.e., $t_{\rm in} \to -\infty$), one finds
\begin{align}
    \chi^{(0)}_{\bm p} = \frac{1}{(2A)^{\frac14}}{\rm e}^{-\frac{\pi}{8}\frac{\mu_{\bm p}^2}{A}}D_{\frac{{\rm i}}{2}\frac{\mu_{\bm p}^2}{A}-\frac12}\left(-{\rm e}^{-\frac{{\rm i}\pi}{4}}\sqrt{2A}(t-t_{\rm prod})\right) \;, \label{eq::2.29}
\end{align}
where $D_\nu(z)$ is the parabolic cylinder function.

Having obtained the mode functions $\chi^{(0)}_{\bm p}$ and $\psi^{(0)}_{\bm k}$, we turn to evaluate the time integral in the production-number formula \eqref{npk}.  Remembering $t_{\rm in} \to -\infty$, we arrive at
\begin{align}
    n_{\bm k}^\psi
        &= \lambda^2 \frac{1}{4 \sqrt{2A}} \int\frac{{\rm d}^3{\bm p}}{(2\pi)^3}\frac{{\rm e}^{-\frac{\pi}{4}\frac{\mu_{\bm p}^2}{A}}}{\Omega_{\bm k}\Omega_{\bm k+\bm p}} \left|\int_{-\infty}^{\infty}{\rm d}t \,{\rm e}^{-{\rm i}(\Omega_{\bm k}+\Omega_{\bm k+\bm p})t}D_{\frac{{\rm i}}{2}\frac{\mu_{\bm p}^2}{A}-\frac12}\left(-{\rm e}^{-\frac{{\rm i}\pi}{4}}\sqrt{2A}(t-t_{\rm prod})\right)\right|^2 \nonumber\\
        &= \lambda^2 \frac{\pi}{(2A)^{\frac{3}{2}}} \int \frac{{\rm d}^3{\bm p}}{(2\pi)^3}\frac{{\rm e}^{-\frac{3\pi}{4}\frac{\mu_{\bm p}^2}{A}}}{\Omega_{\bm k}\Omega_{\bm k+\bm p}}\left|D_{\frac{{\rm i}}{2}\frac{\mu_{\bm p}^2}{A}-\frac12}\left({\rm e}^{{\rm i}\pi/4}\sqrt{\frac{2}{A}}(\Omega_{\bm k}+\Omega_{\bm k+\bm p})\right)\right|^2 \;,\label{nex2} 
\end{align}
where we used $\int_{-\infty}^{\infty}{\rm d}t\, {\rm e}^{-{\rm i}\omega t} D_{{\rm i}\lambda} (-{\rm e}^{{\rm i}\pi/4}t) = 2\sqrt{\pi}{\rm e}^{{\rm i}\pi/4-\pi\lambda/2}D_{+{\rm i}\lambda}(+e^{+{\rm i}\pi/4}2\omega)$ to get the second line.  Note that our evaluation here relies on the Taylor expansion (\ref{wsetup}).  Therefore, the integrand is, strictly speaking, valid only in a finite time region $[-1/\omega, 1/\omega]$, with $\omega$ being the typical frequency of the inflaton $\phi$\footnote{The time-derivatives of the inflaon field $\partial_t^n \phi$ scales as $\phi \propto A \omega^{n-1}$.  Therefore, the ratio of the leading-order term in the Taylor expansion (\ref{wsetup}) to the next-leading one is ${\mathcal O}(\omega(t-t_{\rm prod}))$, which must be small so that the Taylor expansion makes sense.  }.  We have nonetheless approximated the time integration as $[-1/\omega, 1/\omega] \to [-\infty,\infty]$, which is justified since the frequency of the inflaton field $\omega$ is assumed to be smaller than the scales relevant to particle production processes such as the daughter particle energy.  The expression \eqref{nex2} is exact for the model (\ref{wsetup}), though the remaining momentum ${\bm p}$ integration can be carried out either numerically or approximately using, e.g., the saddle-point method.  Note that one may use the saddle-point method to simplify the integrand of Eq.~(\ref{nex2}), which may be useful to understand the rough parameter dependencies of $n_{{\bm k}}^\psi$.  Using an integral representation of the parabolic cylinder function $D_{{\rm i}\nu^2-1/2}(2{\rm e}^{{\rm i}\pi/4}\omega) =\frac{\exp\left[-4{\rm i}\omega^2+\pi\nu^2/4-{\rm i} \pi/8\right]}{\Gamma(-{\rm i}\nu^2+1/2)}\int_{0}^{\infty}{\rm d}y\, y^{-1/2}\ {\rm e}^{-{\rm i} s(y)}$ with $s(y):=y^2/2 + 2\omega y+\nu^2\,{\rm ln}\, y$ and applying the saddle-point method to the $y$ integral, we find
\begin{align}
    \left| D_{{\rm i}\nu^2-\frac{1}{2}}(2e^{\frac{{\rm i}\pi}{4}}\omega) \right|^2
        &\sim \frac{ {\rm e}^{-\frac{\pi\nu^2}{2}}
        \left[ 1 + {\rm e}^{-2\pi\nu^2} \right] }{\sqrt{\nu^2-\omega^2}} \left| \exp\left[ \pi \nu^2 F\left( \frac{\omega}{\nu}\right) \right] \right|^2 \;, \label{saddle}
\end{align}
where
\begin{align}
    F(x) := \frac{-1}{\pi} \left[ x \sqrt{1-x^2}+{\rm i}\,{\rm ln}\left[ x + {\rm i}\sqrt{1-x^2} \right] \right] = \left\{ \begin{array}{ll} {\rm pure\ imaginary} & {\rm for}\ x>1 \\[8pt] \displaystyle \frac{1}{2} - \frac{2}{\pi} x + {\mathcal O}(x^3) & {\rm for}\ x<1 \end{array} \right. \; .
\end{align}
Plugging this expression back into Eq.~(\ref{nex2}), we obtain
\begin{align}
    n_{\bm k}^\psi
    &\sim \lambda^2 \frac{\pi}{4A} \int\frac{{\rm d}^3\bm p}{(2\pi)^3}\frac{ \exp\left(-\pi\frac{\mu_{\bm p}^2}{A}\right) \left[ 1 + \exp\left(-\pi\frac{\mu_{\bm p}^2}{A}\right) \right] }{ \Omega_{\bm k}\Omega_{\bm k+\bm p}\sqrt{\left|\mu_{\bm p}^2-(\Omega_{\bm p}+\Omega_{\bm k+\bm p})^2\right|}} \left| \exp\left[+\frac{\pi}{2} \frac{\mu_{\bm p}^2}{A} F\left( \frac{\Omega_{\bm k}+\Omega_{\bm k+\bm p}}{\mu_{\bm p}} \right)\right] \right|^2 \nonumber\\
    &= \left\{ 
        \begin{array}{l} 
            \begin{array}{l} \displaystyle \lambda^2 \frac{\pi}{4A} \int\frac{{\rm d}^3\bm p}{(2\pi)^3}\frac{ {\rm e}^{-\pi\frac{\mu_{\bm p}^2}{A}} }{ \Omega_{\bm k}\Omega_{\bm k+\bm p}\sqrt{(\Omega_{\bm p}+\Omega_{\bm k+\bm p})^2-\mu_{\bm p}^2}} \left[ 1 + {\mathcal O}\left( {\rm e}^{-\pi\frac{\mu_{\bm p}^2}{A}} \right) \right]  \\[15pt]  \hspace*{100mm} {\rm for}\ \Omega_{\bm k}+\Omega_{\bm k+\bm p} > \mu_{\bm p} \end{array}
            \\[40pt]
            \begin{array}{l} \displaystyle \lambda^2 \frac{\pi}{4A} \int\frac{{\rm d}^3\bm p}{(2\pi)^3}\frac{ {\rm e}^{-\frac{\pi}{2}\frac{\mu_{\bm p}^2}{A}} }{ \Omega_{\bm k}\Omega_{\bm k+\bm p}\sqrt{\mu_{\bm p}^2-(\Omega_{\bm p}+\Omega_{\bm k+\bm p})^2}} \left[ 1 + {\mathcal O}\left( {\rm e}^{-\pi\frac{\mu_{\bm p}^2}{A}}, \frac{\Omega_{\bm k} + \Omega_{\bm k+\bm p}}{\sqrt{A}} \right) \right]  \\[15pt] \hspace*{100mm} {\rm for}\ \Omega_{\bm k}+\Omega_{\bm k+\bm p} < \mu_{\bm p} \end{array}
        \end{array} 
       \right. \;. \label{asym}
\end{align}
This expression is clearly non-perturbative in $A$, so is in the background field $\phi$.  The non-perturbative dependence originates from the dressing of the mode function $\chi^{(0)}_{\bm p}$ by the background $\phi$ field, which is included exactly by the perturbation theory in the Furry picture.

The exponential suppression in the integrand is halved in the regime $\Omega_{\bm k}+\Omega_{\bm k+\bm p} < \mu_{\bm p}$, compared to the other one $\Omega_{\bm k}+\Omega_{\bm k+\bm p} > \mu_{\bm p}$.  This implies a significant amount of particles can be produced from the regime $\Omega_{\bm k}+\Omega_{\bm k+\bm p} < \mu_{\bm p}$ in some parameters.  To clarify this, we solve $\Omega_{\bm k}+\Omega_{\bm k+\bm p} = \mu_{\bm p}$ with respect to $|\bm p|$.  The solution is
\begin{align}
   &|\bm p/a(t_{\rm prod})| \nonumber\\
        &= \frac{m_\chi^2}{2(k^2\sin^2\theta/a^2(t_{\rm prod})+m_\psi^2)}\left[\pm \Omega_{\bm k}\sqrt{1-\frac{4(k^2\sin^2\theta/a^2(t_{\rm prod})+m_\psi^2)}{m_\chi^2}}-|\bm k|\cos\theta \right] \nonumber\\
        &\approx \frac{m_\chi^2}{2m_\psi}\sqrt{1-\frac{4m_\psi^2}{m_\chi^2}} \;, \label{eq:2.35}
\end{align}
where $\cos\theta=\frac{\bm k \cdot\bm p}{|\bm k||\bm p|}$ and we take the minus sign if $\cos\theta<0$ so that the right-hand side is positive.  For simplicity, we considered a small $\bm k$ limit such that $|\bm k|<m_\psi$ in the second line.  Thus, the regime $\Omega_{\bm k}+\Omega_{\bm p+\bm k} < \mu_{\bm p}$ ($\Omega_{\bm k}+\Omega_{\bm p+\bm k} > \mu_{\bm p}$) contributes to the integral when $|{\bm p}/a(t_{\rm prod})|$ is below (above) the right-hand side of Eq.~(\ref{eq:2.35}).  Importantly, the regime $\Omega_{\bm k}+\Omega_{\bm p+\bm k} > \mu_{\bm p}$ can contribute only when ${\bm p}$ is large.  Such large ${\bm p}$ contributions are, however, strongly suppressed, as the integrand in Eq.~\eqref{asym} has an exponential factor depending on $|\bm p|^2$, or $\frac{\pi|\bm p|^2}{Aa^2(t_{\rm prod})} \gg 1$.  Thus, in the limit of $m_\chi/m_\psi \to \infty$, i.e., if the daughter particle mass $m_\psi$ is much smaller than the bare mass of the parent particle $m_\chi$, the regime $\Omega_{\bm k}+\Omega_{\bm p+\bm k} > \mu_{\bm p}$ can give only a negligible contribution, and the integral is dominated by the contribution from the other regime $\Omega_{\bm k}+\Omega_{\bm p+\bm k} < \mu_{\bm p}$.  On the other hand, if $m_\psi/m_\chi$ is not very large, the contribution from the regime $\Omega_{\bm k}+\Omega_{\bm p+\bm k} > \mu_{\bm p}$ is not necessarily negligible, and hence one needs to take into account both contributions properly.

We emphasize that in the non-perturbative result (\ref{nex2}), or its asymptotic form (\ref{asym}), the daughter $\psi$ particles can be produced with any momenta ${\bm k}$ and masses $m_\chi, m_\psi$ due to the non-perturbative dressing by the background $\phi$ field.  As we clarify in the following subsections, this is in contrast to the phenomenological perturbative treatment, in which the decay $\ket{\chi} \to \ket{\psi\psi}$ is restricted by a kinematical condition $\omega_{\bm p} = \Omega_{\bm k} + \Omega_{{\bm k}+{\bm p}}$, and consequently the production of the daughter $\psi$ particles can be prohibited for some values of ${\bm k}$, $m_\chi$, and $m_\psi$.  Intuitively, in the non-perturbative case $\chi$ particles are scattered by the background $\phi$ field infinite times during the decay process which modifies the kinematical condition by supplying energy, while such a modification is absent in the perturbative case.

To compare the non-perturbative result (\ref{nex2}) with the phenomenological perturbative calculation in the next subsection, it is convenient to decompose the number of $\psi$ particles,
\begin{align}
    n_{\bm k}^\psi = 2\int \frac{{\rm d}^3{\bm p}}{(2\pi)^3} \, n_{\bm p}^\chi \, P^{\rm np}_{{\bm k},{\bm p}}(\chi \to \psi\psi) \; , \label{eq::2.31}
\end{align}
to estimate the decay probability of $\chi$ particles $P^{\rm np}_{{\bm k},{\bm p}}(\chi \to \psi\psi)$ in the presence of the slow background $\phi$ field.  Note that the number 2 is inserted in Eq.~(\ref{eq::2.31}) because two $\psi$ particles are produced per a decay $\ket{\chi} \to \ket{\psi\psi}$.  The number of parent $\chi$ particles produced from the vacuum can be obtained straightforwardly by computing the Bogoliubov coefficient $\beta_{\bm p}$ (\ref{eq:2.20}) from the dressed mode function $\chi^{(0)}_{\bm p}$ (\ref{eq::2.29}) (see e.g. \cite{Kofman:1997yn}), which yields 
\begin{align}
    n_{\bm p}^\chi = {\rm e}^{- \pi \frac{\mu_{\bm p}^2}{A}} \left[ 1 + {\mathcal O}(\lambda) \right] \; .  \label{eq::2.32}
\end{align}
Thus, the decay probability can be estimated as
\begin{align}
    P^{\rm np}_{{\bm k},{\bm p}}(\chi \to \psi\psi) = \lambda^2\frac{\pi}{2 (2A)^{\frac32}} \frac{ {\rm e}^{+\frac{\pi}{4}\frac{\mu_{\bm p}^2}{A}}}{\Omega_{\bm k}\Omega_{\bm k+\bm p}}\left|D_{\frac{{\rm i}}{2}\frac{\mu_{\bm p}^2}{A}-\frac12}\left({\rm e}^{{\rm i}\pi/4}\sqrt{\frac{2}{A}}(\Omega_{\bm k}+\Omega_{\bm k+\bm p})\right)\right|^2 \label{eq::2.33}
\end{align}
up to ${\mathcal O}(\lambda^3)$.

\subsection{Phenomenological perturbative approach} \label{sec2.4}

We analytically compare the non-perturbative formula (\ref{nex2}) with the phenomenological perturbative approach in the slow limit to clarify differences between the two.  In particular, we shall argue that the phenomenological perturbative approach is valid in a weak-field limit such that the inflaton amplitude is not large and that there exists a parameter regime (dubbed as ``kinematically-forbidden'' parameter regime) such that the phenomenological perturbative approach cannot access.  

The idea of the phenomenological perturbative approach is to estimate the number of daughter $\psi$ particles using the decomposition (\ref{eq::2.31}), i.e., compute a convolution integral between the number of parent $\chi$ particles $n_{\bm p}^\chi$ and a {\it perturbative} decay probability $P^{\rm pert}_{{\bm k},{\bm p}}(\chi \to \psi\psi)$.  This is essentially the same idea that the original work on instant preheating \cite{Felder:1998vq} proposed.  Nonetheless, what we describe below can be understood as a sort of generalizations that includes momentum dependencies and the time-evolution history of the background $\phi$ field.  

The number of parent $\chi$ particles can naturally be estimated with the Bogoliubov coefficient $\beta_{\bm p}$ as $n_{\bm p}^\chi = |\beta_{\bm p}|^2 [1+{\mathcal O}(\lambda)]$, or by the exponential formula (\ref{eq::2.32}) in the slow-field limit.  Thus, the main task is to compute the perturbative decay probability, which is the origin of the difference between the non-perturbative formula (\ref{nex2}) and the phenomenological perturbative approach as we describe below.  

To obtain the perturbative decay probability $P^{\rm pert}_{{\bm k},{\bm p}}(\chi \to \psi\psi)$, we first compute the corresponding decay probability without any background in the flat spacetime.  At the tree level ${\mathcal O}(\lambda^2)$, it reads 
\begin{align}
    P^{\rm tree}_{{\bm k},{\bm p}}(\chi \to \psi\psi) 
    &= \left| 
    \begin{tikzpicture}[baseline={([yshift=-.5ex]current bounding box.center)},vertex/.style={anchor=base,
    circle,fill=black!25,minimum size=18pt,inner sep=2pt}]
	\begin{feynhand}
	\vertex[particle] (c1) at (-2,0) {$\chi$}; 
	\vertex[particle] (v) at (0,0); 
	\vertex[particle] (p1) at (1.5,-1) {$\psi$}; 
	\vertex[particle] (p2) at (1.5,+1) {$\psi$}; 
	\propag[fermion] (c1) to [edge label = ${\bm p}$] (v);
	\propag[chasca] (v) to [edge label = $-{\bm k}+{\bm p}$] (p1);
	\propag[chasca] (v) to [edge label = ${\bm k}$] (p2);
	\end{feynhand}
	\end{tikzpicture} 
	\right|^2 \nonumber\\
    &= \left|\lambda \int^{\infty}_{-\infty}{\rm d}t' \chi^{{\rm un}*}_{\bm p}(t')\psi^{\rm un}_{-{\bm k}+{\bm p}}(t')\psi^{\rm un}_{{\bm k}}(t')\right|^2  \nonumber\\
    &= \lambda^2 \frac{\pi}{8}T \frac{1}{\omega^{\rm un}_{\bm p}\Omega^{\rm un}_{{\bm k}}\Omega^{\rm un}_{-{\bm k}+{\bm p}}} \delta(\omega^{\rm un}_{\bm p}-\Omega^{\rm un}_{{\bm k}}-\Omega^{\rm un}_{-{\bm k}+{\bm p}}) \;, \label{eq::2.34}
\end{align}
where $T$ is the whole time-interval and $\chi^{\rm un}_{\bm p} = \frac{{\rm e}^{-{\rm i}\omega^{\rm un}_{\bm p}t}}{\sqrt{2\omega^{\rm un}_{\bm p}}}$ and $\psi^{\rm un}_{\bm k} = \frac{{\rm e}^{-{\rm i}\Omega^{\rm un}_{\bm k}t}}{\sqrt{2\Omega^{\rm un}_{\bm k}}}$ are the un-dressed mode functions (i.e., plane waves) for $\chi$ and $\psi$ particles with frequencies $\omega^{\rm un}_{\bm p}$ and $\Omega^{\rm un}_{\bm k}$, respectively.

The perturbative decay probability is then obtained rather phenomenologically by incorporating background effects to the tree-level decay probability  $P^{\rm tree}_{{\bm k},{\bm p}}(\chi \to \psi\psi)$ (\ref{eq::2.34}) with an ad-hoc prescription: Replace all the un-dressed quantities with the dressed ones, i.e., $\omega^{\rm un}_{\bm p}, \Omega^{\rm un}_{\bm k} \to \omega_{\bm p}(t), \Omega_{\bm k}(t)$ and ${\bm p},{\bm k} \to {\bm p}/a, {\bm k}/a$.  The time factor $T$ is also replaced with $\int {\rm d}t$, so that all the time dependencies appeared by the replacement are integrated out.\footnote{A similar procedure, known as ``locally-constant field approximation (LCFA)'' has been used to calculate probabilities of QED processes under coherent electromagnetic fields \cite{Ritus1985}.  }  The resulting expression is 
\begin{align}
    P^{\rm tree}_{{\bm k},{\bm p}}(\chi \to \psi\psi) 
    &\to \lambda^2 \frac{\pi }{8} \int {\rm d}t\,\frac{1}{ \omega_{\bm p}(t)\Omega_{{\bm k}}(t)\Omega_{-{\bm k}+{\bm p}}(t)} \delta(\omega_{\bm p}(t)-\Omega_{{\bm k}}(t)-\Omega_{-{\bm k}+{\bm p}}(t)) \nonumber\\
    &=: P^{\rm pert}_{{\bm k},{\bm p}}(\chi \to \psi\psi) \;. \label{eq::2.35}
\end{align}

Importantly, the perturbative decay probability $P^{\rm pert}_{{\bm k},{\bm p}}(\chi \to \psi\psi)$ (\ref{eq::2.35}) has a delta-function factor, which accounts for the kinematics of the decay within the un-dressed particle states.  It indicates that, in the phenomenological perturbative approach, the decay occurs only if there exists an instant of time such that the energies of the parent $\chi$ particle and the daughter two $\psi$ particles match $\omega_{\bm p}(t)=\Omega_{\bm k}(t)+\Omega_{-\bm k+\bm p}(t)$.  This is in contrast to the non-perturbative formula (\ref{eq::2.33}): there is no such thresholds due to the non-perturbative scattering between the background $\phi$ field and the parent $\chi$ field (i.e., the dressing of the mode function $\chi^{(0)}_{\bm p}$), by which the parent $\chi$ particles acquire additional energy and thus the kinematics is modified.  Such a scattering effect is completely dismissed in the ad-hoc replacement of the phenomenological perturbative approach $\omega^{\rm un}_{\bm p}, \Omega^{\rm un}_{\bm k} \to \omega_{\bm p}(t), \Omega_{\bm k}(t)$.

Putting everything together, one arrives at the phenomenological perturbative formula for the number of daughter $\psi$ particles: 
\begin{align}
    n_{\bm k}^\psi 
    &= 2\int \frac{{\rm d}^3{\bm p}}{(2\pi)^3} \, n_{\bm p}^\chi \, P^{\rm pert}_{{\bm k},{\bm p}}(\chi \to \psi\psi) \nonumber\\
    &= \lambda^2 |\beta_{\bm p}|^2 \frac{\pi}{4} \int {\rm d}t\int\frac{{\rm d}^3{\bm p}}{(2\pi)^3}\frac{1}{\Omega_{\bm k}(t)\Omega_{-\bm k+\bm p}(t)\omega_{\bm p}(t)} \delta\left( \omega_{\bm p}(t)-\Omega_{\bm k}(t)-\Omega_{-\bm k+\bm p}(t) \right) \nonumber\\
    &= \lambda^2 |\beta_{\bm p}|^2 \frac{\pi}{4} \int {\rm d}t\int\frac{{\rm d}^3{\bm p}}{(2\pi)^3}\frac{1}{\Omega_{\bm k}(t)\Omega_{\bm k+\bm p}(t)\omega_{\bm p}(t)} \delta\left( \omega_{\bm p}(t)-\Omega_{\bm k}(t)-\Omega_{\bm k+\bm p}(t) \right) \; .\label{eq::2.36}
\end{align}
In the last line, we flipped the sign of ${\bm p} \to -{\bm p}$ and used $\beta_{-\bm p} = \beta_{{\bm p}}$, $\omega_{-{\bm p}} = \omega_{\bm p}$, and $\Omega_{-{\bm k}-{\bm p}} = \Omega_{{\bm k}+{\bm p}}$ (which are the manifestations of the parity invariance).  Note that Eq.~(\ref{eq::2.36}) includes all the contributions from the time interval $\int {\rm d}t$, while Ref.~\cite{Felder:1998vq} focuses on those from particular times at which the background $\phi$ field takes its (local) maxima (as the particle production may presumably be dominated by such times but this is not necessarily true, e.g., if the background $\phi$ field is sufficiently slow and sizable number of particles can be produced before reaching the maxima).  In the slow-field limit~(\ref{wsetup}), one may further simplify Eq.~(\ref{eq::2.36}).  Noting
\begin{align}
    \delta\left(\omega_{\bm p}-\Omega_{\bm k}-\Omega_{\bm k+\bm p}\right)
    &=\frac{\Omega_{\bm k}+\Omega_{\bm k+\bm p}}{A\sqrt{(\Omega_{\bm k}+\Omega_{\bm k+\bm p})^2-\mu_{\bm p}^2}} \Theta\left( (\Omega_{\bm k}+\Omega_{\bm k+\bm p})^2-\mu_{\bm p}^2 \right) \nonumber\\
    &\quad \times \delta\left( t- \frac{\sqrt{(\Omega_{\bm k}+\Omega_{\bm k+\bm p})^2-\mu_{\bm p}^2}}{A}  \right)\;,
\end{align}
one may explicitly carry out the time integration to find
\begin{align}
    n_{\bm k}^\psi = \lambda^2 \frac{\pi}{4A} \int\frac{{\rm d}^3{\bm p}}{(2\pi)^3}\frac{{\rm e}^{-\pi\frac{\mu_{\bm p}^2}{A}}}{\Omega_{\bm k}\Omega_{\bm k+\bm p}\sqrt{(\Omega_{\bm k}+\Omega_{\bm k+\bm p})^2-\mu_{\bm p}^2}} \Theta\left( (\Omega_{\bm k}+\Omega_{\bm k+\bm p})^2-\mu_{\bm p}^2 \right) \;. \label{nLCFA}
\end{align}

The step function in the phenomenological perturbative formula (\ref{nLCFA}) defines the ``kinematically-allowed'' parameter regime for the daughter $\psi$ particles production.  In our model (\ref{wsetup}), the parent particle $\chi$ is produced at the moment $t=t_{\rm prod}$ and since then the mass of $\chi$, or the frequency $\omega_{\bm p}$, monotonically increases in time.  The phenomenological perturbative formula based on the un-dressed mode functions (\ref{eq::2.36}) thus indicates that the daughter particles must be heavy or have sufficiently large momentum to satisfy the kinematic constraint $\Omega_{\bm k}+\Omega_{\bm k+\bm p} = \omega_{\bm p}(t) \geq \mu_{\bm p}$ at some time $t>t_{\rm prod}$ so that they can be produced.  Conversely, daughter $\psi$ particles can never be produced outside of the kinematically-allowed parameter regime (i.e., $\Omega_{\bm k}+\Omega_{\bm k+\bm p} < \mu_{\bm p}$) within the perturbative formula (\ref{eq::2.36}), whereas there is no such prohibition in the non-perturbative formula (\ref{nex2}) due to the change of the kinematics by the dressing.  We shall discuss more about this particle production in the kinematically-forbidden parameter regime in the next subsection~\ref{sec2.5}.

The phenomenological perturbative formula (\ref{nLCFA}) is valid only in the kinematically-allowed regime and only if the backgrounds are weak.  In such a situation, the ad-hoc replacements to obtain the perturbative decay rate [i.e., $\omega^{\rm un}_{\bm p}, \Omega^{\rm un}_{\bm k} \to \omega_{\bm p}(t), \Omega_{\bm k}(t)$, ${\bm p},{\bm k} \to {\bm p}/a, {\bm k}/a$, and $T \to \int {\rm d}t$] may work as the system is approximately invariant under the time translation and hence the plane waves may give good approximations to the dressed mode functions.  In other words, if the parameter $A$ in the slow-field expansion (\ref{wsetup}), which controls the magnitude of the inflaton oscillation is very small, the dressed mode functions may reduce to the plane waves and then the phenomenological perturbative approach (\ref{eq::2.35}) may work.  Indeed, the asymptotic form of the non-perturbative result (\ref{asym}) precisely coincides with the perturbative one (\ref{nLCFA}) for $(\Omega_{\bm k}+\Omega_{\bm k+\bm p})^2-\mu^2_{\bm p} > 0$ (i.e., in the kinematically-allowed regime) after taking the limit of $A \to 0$ to drop the error term ${\mathcal O}({\rm e}^{-\frac{\pi \mu_{\bm p}}{A}}) \to 0$.  One may also confirm the consistency between the two in another way: by using the asymptotic form of the parabolic cylinder function $D_\nu(z) \xrightarrow{|z| \to \infty} {\rm e}^{-z^2/4}z^\nu$, one may directly take the limit of $A \to 0$, or $\sqrt{\frac{2}{A}}(\Omega_{\bm k}+\Omega_{\bm k+\bm p}) \to \infty$, in Eq.~(\ref{nex2}) to find
\begin{align}
    n_{\bm k}^\psi \xrightarrow{A \to 0} \lambda^2 \frac{\pi}{4A} \int\frac{{\rm d}^3\bm p}{(2\pi)^3}\frac{ {\rm e}^{-\pi \frac{\mu_{\bm p}^2}{A}}}{\Omega_{\bm k}\Omega_{\bm k+\bm p}(\Omega_{\bm k}+\Omega_{\bm k+\bm p})} \; .  \label{asymptotic}
\end{align}
This reproduces Eq.~(\ref{nLCFA}) provided that $(\Omega_{\bm k}+\Omega_{\bm k+\bm p})^2-\mu^2_{\bm p} > 0$ and that $\sqrt{(\Omega_{\bm k}+\Omega_{\bm k+\bm p})^2-\mu_{\bm p}^2} \sim \Omega_{\bm k}+\Omega_{\bm k+\bm p}$.

\subsection{Kinematically-forbidden particle production} \label{sec2.5}

\begin{figure}[t]
\begin{center}
\includegraphics[clip, width=0.495\textwidth]{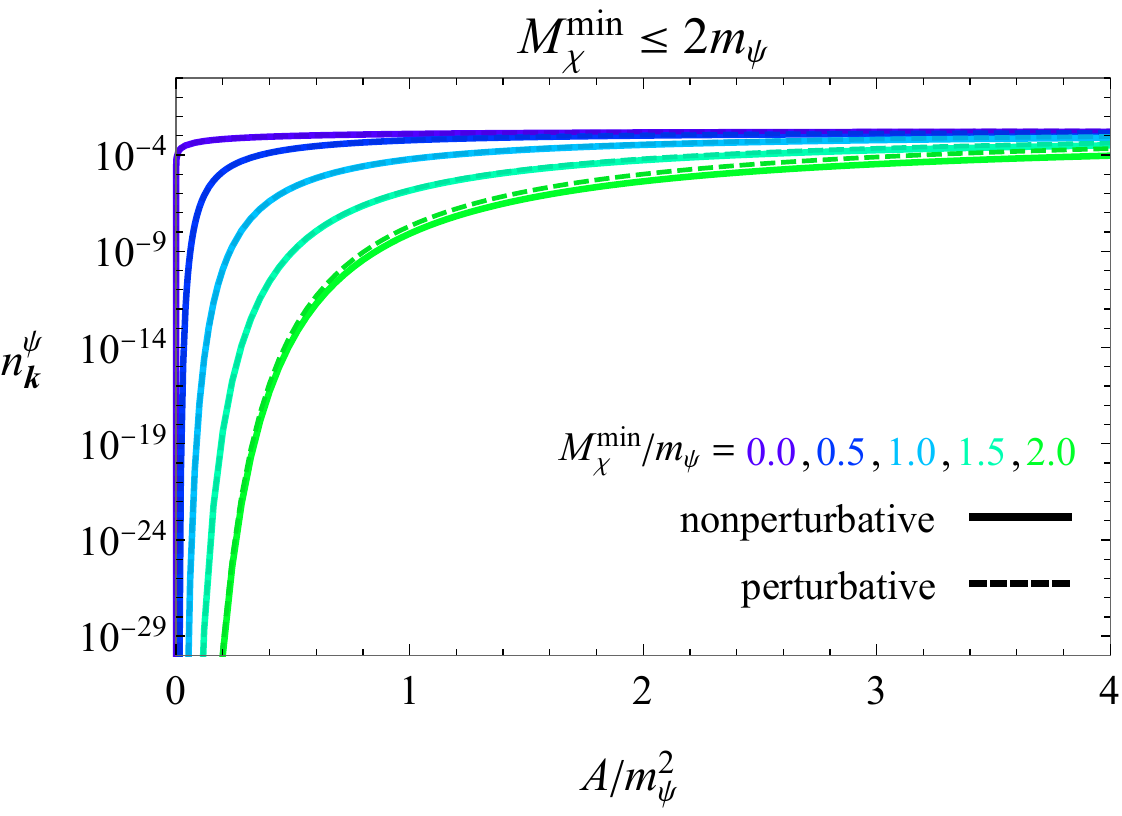} 
\includegraphics[clip, width=0.495\textwidth]{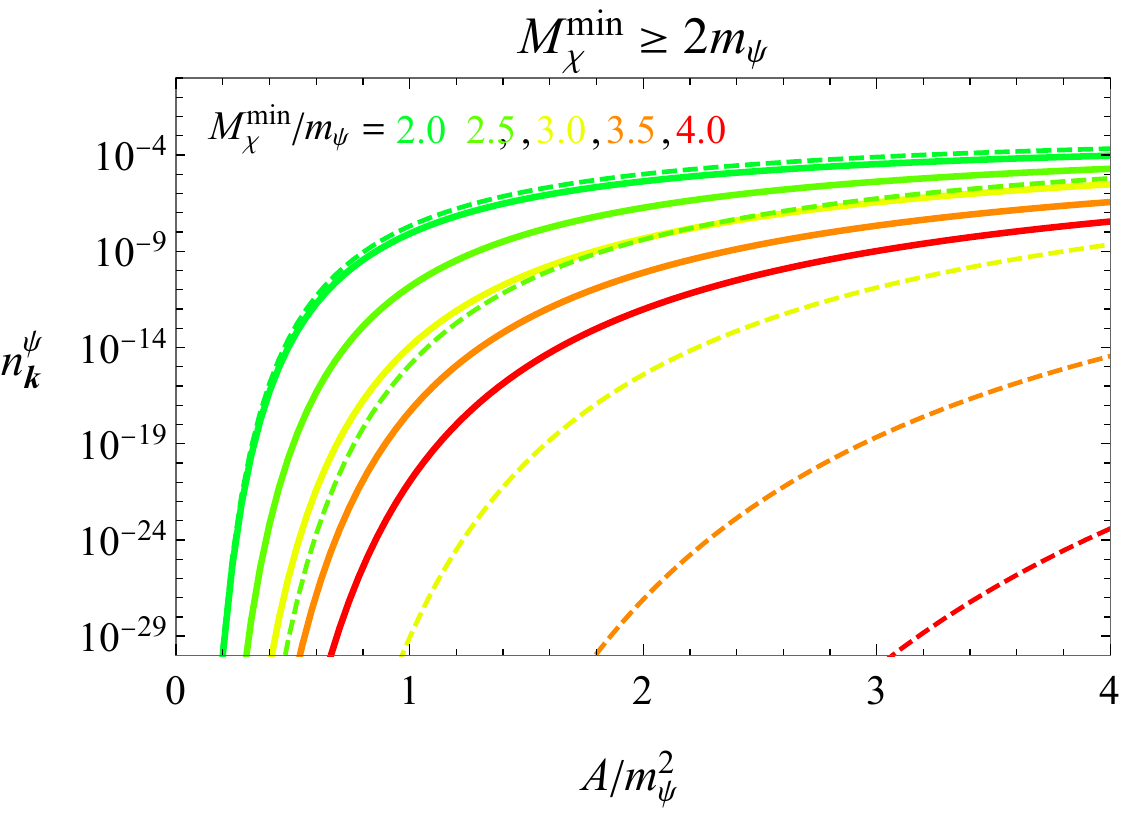}  
\end{center}
\caption{\label{fig2} The number of daughter particles produced $n^{\psi}_{\bm k}$ at ${\bm k} = {\bm 0}$ as a function of the parameter $A$ with various values of the minimum of the parent particle mass $M^{\rm min}_\chi = \sqrt{m_\chi^2 + \zeta \phi^2(t_{\rm prod})}$.  The thick and dashed lines show the non-perturbative result in the slow-field limit (\ref{eq::2.33}) and the phenomenological perturbative result (\ref{eq::2.36}), respectively.  The other parameters are fixed as $a(t_{\rm prod})=1$ and $\lambda=1$.  }
\end{figure}

\begin{figure}[t]
\begin{center}
\includegraphics[clip, width=0.495\textwidth]{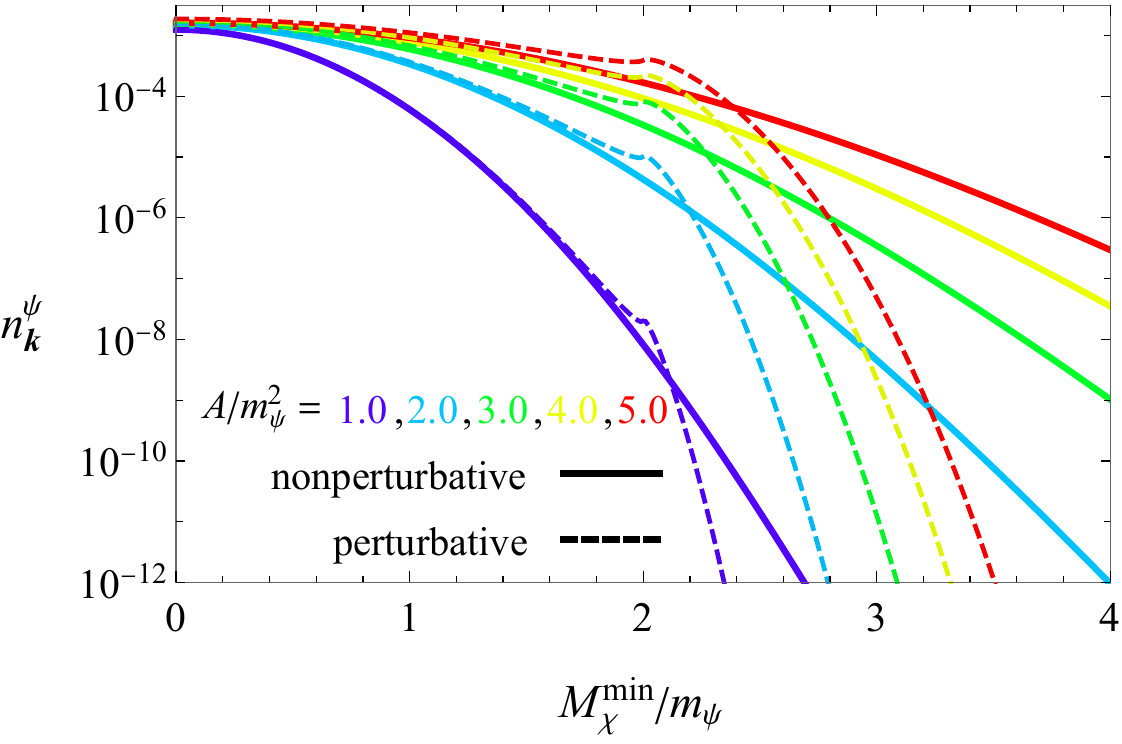} 
\end{center}
\caption{\label{fig3} The number of daughter particles produced $n^{\psi}_{\bm k}$ at ${\bm k} = {\bm 0}$ as a function of the minimum of the parent particle mass $M^{\rm min}_\chi = \sqrt{m_\chi^2 + \zeta \phi^2(t_{\rm prod})}$ for various values of $A$.  The thick and dashed lines show the non-perturbative result in the slow-field limit (\ref{eq::2.33}) and the phenomenological perturbative result (\ref{eq::2.36}), respectively.  The other parameters are fixed as $a(t_{\rm prod})=1$ and $\lambda=1$.  }
\end{figure}

\begin{figure}[t]
\begin{center}
\includegraphics[clip, width=0.495\textwidth]{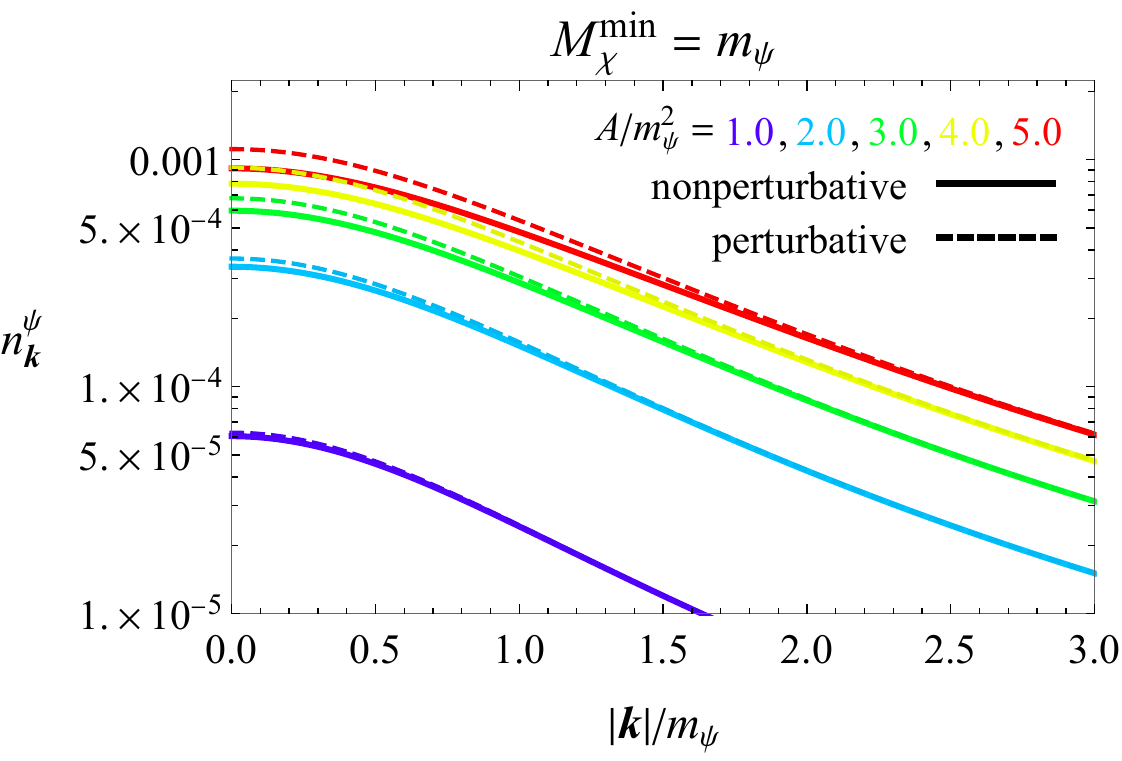} 
\includegraphics[clip, width=0.495\textwidth]{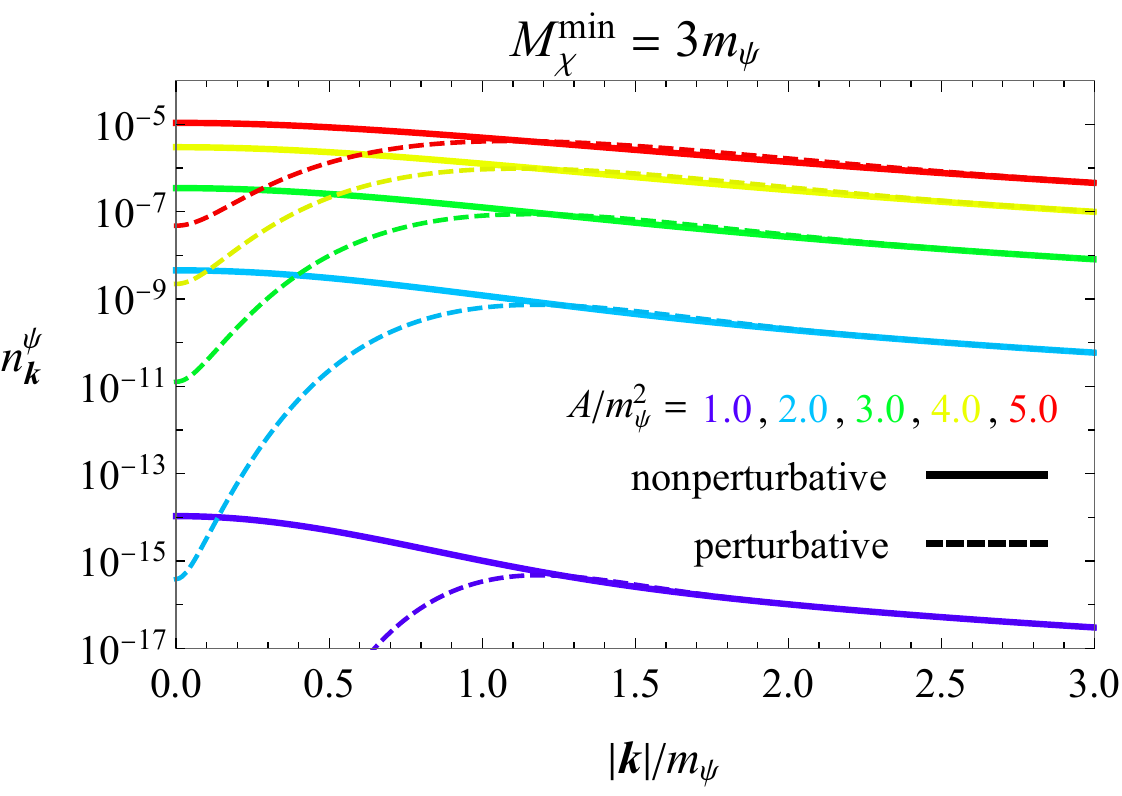}  
\end{center}
\caption{\label{fig4} The number of daughter particles produced $n^{\psi}_{\bm k}$ as a function of the momentum ${\bm k}$ for $M^{\rm min}_\chi = m_\psi$ (left) and $M^{\rm min}_\chi = 3m_\psi$ (right) with various values of $A$.  The thick and dashed lines show the non-perturbative result in the slow-field limit (\ref{eq::2.33}) and the phenomenological perturbative result (\ref{eq::2.36}), respectively.  The other parameters are fixed as $a(t_{\rm prod})=1$ and $\lambda=1$.  }
\end{figure}

We have shown in the previous subsections that the non-perturbative formula (\ref{nex2}) is free from the kinematic constraint of the phenomenological perturbative approach, and hence daughter $\psi$ particles can be produced even in the kinematically-forbidden parameter regime.  In this subsection, we take a close look at this kinematically-forbidden particle production and show that it is crucially important when daughter $\psi$ particles are light and/or parent $\chi$ particles are heavy.  

We first discuss whether the kinematically-forbidden process can be understood within the perturbative decay picture. The naively-defined decay probability~\eqref{eq::2.33} in the kinematically-forbidden process is 
\begin{align}
    P^{\rm np}_{{\bm k},{\bm p}}(\chi \to \psi\psi) \approx \lambda^2 \frac{\pi}{8A} \int\frac{{\rm d}^3\bm p}{(2\pi)^3}\frac{ {\rm e}^{+\frac{\pi}{2}\frac{\mu_{\bm p}^2}{A}} }{ \Omega_{\bm k}\Omega_{\bm k+\bm p}\sqrt{\mu_{\bm p}^2-(\Omega_{\bm p}+\Omega_{\bm k+\bm p})^2}}.\label{Pforbidden}
\end{align}
This quantity becomes exponentially large when $\frac{\pi}{2}\frac{\mu_{\bm p}^2}{A}>1$, which would violate the unitarity (i.e., the probability exceeds unity) if we literally interpreted it as the decay probability of $|\chi\rangle\to|\psi\psi\rangle$. What is wrong here would be the naive perturbative-decay interpretation of the kinematically-forbidden process. A more suitable interpretation is that the kinematically-forbidden process is caused by multiple scatterings with the inflaton field, which is fairly non-perturbative in the coupling to the inflaton. (See also Fig.~\ref{fig1}.)  As discussed below Eq.~\eqref{asym}, for light daughter- and/or heavy parent-particles, the kinematically-forbidden particle production, which cannot be captured by the phenomenological perturbative approach, gives the dominant contribution.  Therefore the kinematically-forbidden particle production must be included to correctly estimate the number of daughter particles produced in a certain parameter regime.  

We numerically integrate the integrand $n_{{\bm k},{\bm p}}^\psi$ over the momentum ${\bm p}$ to discuss the behavior of $n_{\bm k}^\psi$ in various parameter ranges and the role of the kinematically-forbidden particle production.  As shown in Figs.~\ref{fig2} and \ref{fig3}, the phenomenological perturbative formula~\eqref{eq::2.36} coincides with the non-perturbative result (\ref{nex2}) when the daughter/parent particle is sufficiently heavy/light such that $M^{\rm min}_\chi < 2m_\psi$, for which the kinematic condition is satisfied for any momenta ${\bm p}$ and ${\bm k}$.  The coincidence looks almost insensitive to the values of $A$ considered here (though there appear small deviations for larger values of $A$ and the deviation becomes larger for larger $A$).  On the other hand, when the daughter/parent particle gets light/heavy $M^{\rm min}_\chi > 2m_\psi$, the phenomenological perturbative formula~\eqref{eq::2.36} completely fails to reproduce the non-perturbative result (\ref{nex2}).  This is because the kinematic condition is not necessarily satisfied for $M^{\rm min}_\chi > 2m_\psi$, and hence the kinematically-forbidden particle production becomes crucial.  As seen from Fig.~\ref{fig4}, the kinematically-forbidden particle production is more manifest in the low-momentum region, while high-momentum modes are consistent with the perturbative results as the kinematic condition is satisfied for such modes.

\section{Application to a realistic model}\label{sec3}

On the basis of the discussions so far, we consider a more realistic model of preheating and discuss implications of the kinematically-forbidden particle production.  We model the classical backgrounds as~\cite{Kofman:1997yn}
\begin{align}
    a(t) = \left( \frac{m_\phi t}{\pi} \right)^{\frac{2}{3}} \;,\quad \phi(t)=\frac{C}{t}\sin m_\phi t \;,\label{phsetup}
\end{align}
where $m_\phi$ denotes the mass of inflaton $\phi$ and $C\sim 2\sqrt{6}M_{\rm pl}/3m_\phi \gg 1$, with $M_{\rm pl} \sim 2.4 \times 10^{18}\;{\rm GeV}$ being the (reduced) Planck mass, is the initial amplitude of the inflaton oscillation.  The scale factor is normalized so that $a(\pi m_\phi^{-1})=1$ at which time the inflaton field $\phi(t)$ first crosses the zero $0 = \phi(\pi m_\phi^{-1})$.  This setup describes the coherent oscillation of inflaton after the end of inflation, where it behaves as non-relativistic matter component, and therefore $a(t)\propto t^{2/3}$ according to the Friedmann equation.  The amplitude of the inflaton $\phi$ decays in time due to the expansion, and hence only the first few oscillations can be important.  We also assume that $\chi$ particle is sufficiently heavy so that the $\chi$ production is dominated only by the first oscillation as the parent $\chi$ particle production is exponentially suppressed by the ratio of $m_\chi$ to the amplitude of the inflaton oscillation [see Eq.~(\ref{eq:3.5})].  More precisely, we assume a scale hierarchy $\sqrt{\zeta}C m_\phi^2 \sim \sqrt{\zeta} m_\phi M_{\rm pl} \lesssim m_{\chi}^2  \ll \zeta C^2 m_\phi^2 \sim \zeta M_{\rm pl}^2$.  Note that the first inequality is ``$\lesssim$'' here, not ``$\ll$,'' and our discussion below includes either case of $m_\chi^2/\sqrt{\zeta}C m_\phi^2 \sim \mathcal{O}(1)$ or $m_\chi^2/\sqrt{\zeta}C m_\phi^2\gg 1$.  We shall not consider light $\chi$ particles.  In such a case, the $\chi$ production becomes significant in the early phase, which is called the broad resonance regime~\cite{Kofman:1997yn}.  There, the backreaction of the $\chi$-particle production to the inflaton $\phi$ is no longer negligible, which may be dealt with, e.g., lattice simulations~\cite{Khlebnikov:1996mc,Khlebnikov:1996wr,Prokopec:1996rr,Khlebnikov:1996zt}. We note that the parameter regime we consider in the following differs from the original work of instant preheating~\cite{Felder:1998vq}, where the authors considered the case of light $\chi$-particles $m_\chi^2\ll \sqrt\zeta Cm_\phi^2$ and the daughter particles are fermions.  In this parameter regime, the parent particle $\chi$ can be produced more efficiently, for which one has to take into account multiple vacuum pair production events that occur around $\phi(t) = 0$.  In our parameter regime, only the first pair creation contributes, and we can safely neglect the later events.  This simplifies the analysis of scatterings between parent and daughter particles. 
Note that the properties of the vacuum pair production of the parent $\chi$ particles is essentially unaffected by the mass $m_\chi$.  The mass $m_\chi$ only affects the probability through the exponential formula (\ref{eq::2.32}), and the the production time is unchanged and takes place at $\phi(t) = 0$ (see e.g. Sec.~4.3 of \cite{Yamada:2021kqw} and footnote~\ref{foot3}).  

\begin{figure}[htbp]
\centering
\includegraphics[width=.6\textwidth]{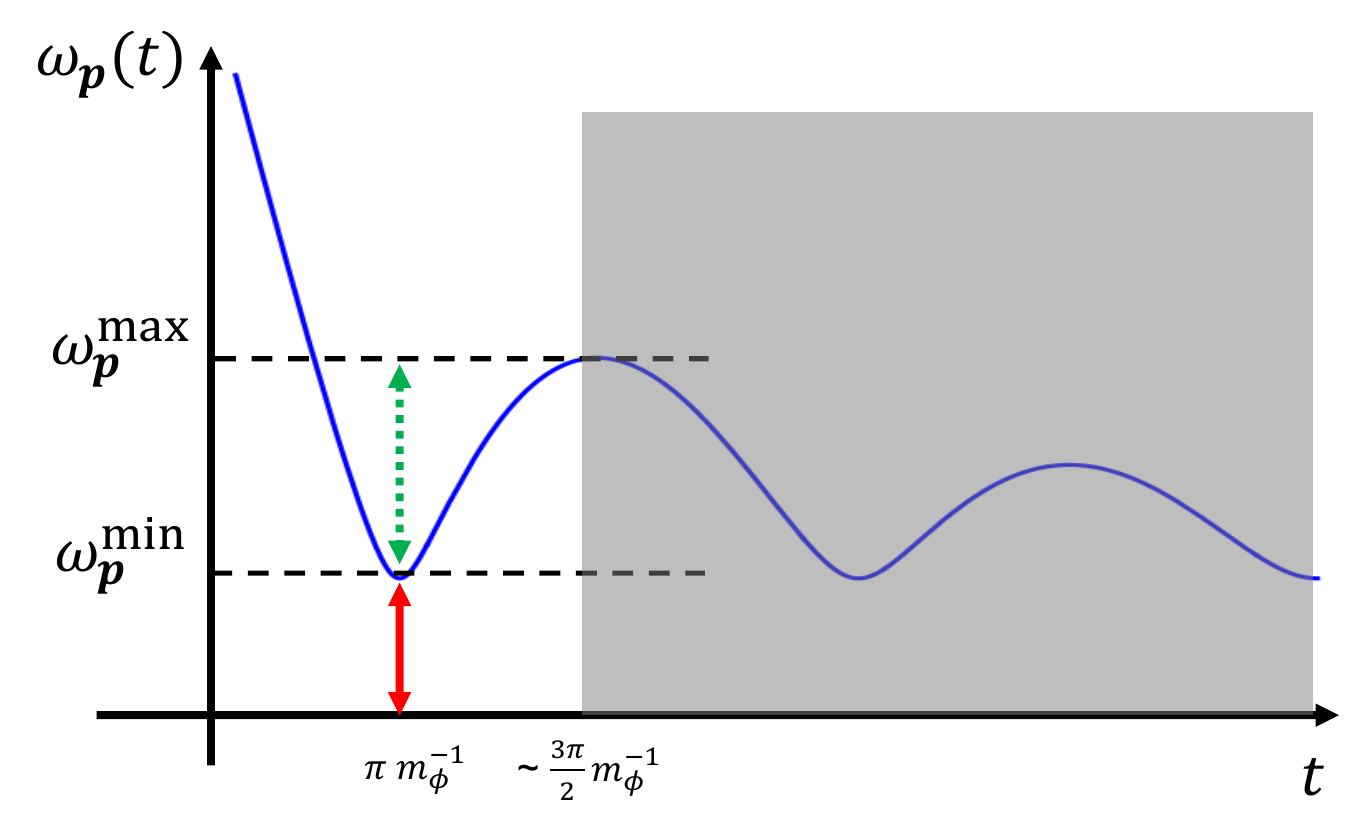}
\caption{A schematic figure of the effective frequency of $\chi$ particle $\omega_{\bm p}(t)$.  The full effective frequency shows a dumped oscillation, while we focus only around the first minimum $m_\phi t \sim \pi$ (i.e., neglect the shaded region), where the dominant number of parent $\chi$ particles are presumably produced.  The perturbative decay $\ket{\chi}\to\ket{\psi\psi}$ is possible if $\omega_{\bm p}^{\rm min}<\Omega_{\bm k}+\Omega_{\bm k+\bm p}<\omega^{\rm max}_{\bm p}$ (namely, if $\Omega_{\bm k}+\Omega_{\bm k+\bm p}$ is within the range indicated by a green dotted line).  If $\Omega_{\bm k}+\Omega_{\bm k+\bm p} < \omega_{\bm p}^{\rm min}$ (i.e. $\Omega_{\bm k}+\Omega_{\bm k+\bm p}$ is within the range indicated by a red line), the kinematically-forbidden $\psi$ production occurs.   }\label{fig;omegaeff}
\end{figure}

To apply the result of Sec.~\ref{sec:2}, we expand the time-dependent frequencies $\omega_{\bm p}$ and $\Omega_{\bm k}$ for $\chi$ and $\psi$ particles, respectively, by assuming that the inflaton field $\phi$ and the expansion $a(t)$ are sufficiently slow.  As we have assumed that the bare mass of the parent particle $m_\chi$ is sufficiently large $m_{\chi}^2 \gtrsim \sqrt{\zeta}C m_\phi^2$, it is reasonable to take account only of the first oscillation, which dominates the production events.  Thus, one may identify the first minimum as the production time $t_{\rm prod}=\pi m_\phi^{-1}$ (see Fig.~\ref{fig;omegaeff}) and finds 
\begin{align}
\begin{split}
    \omega_{\bm p}^2(t) 
        &= \frac{{\bm p}^2}{a^2(t)} + m_\chi^2 + \frac{\zeta C^2}{t^2}\sin^2(m_\phi t)
        \sim {\bm p}^2 + m_\chi^2 + \frac{\zeta C^2m_\phi^4}{\pi^2}(t-\pi m_\phi^{-1})^2 \;, \\
    \Omega_{\bm k}^2(t) 
        &= \frac{{\bm k}^2}{a^2(t)} + m_\psi^2 
        \sim {\bm k}^2 + m_\psi^2 \; . \label{eq:3.2}
\end{split}
\end{align}
Note that $\omega_{\bm p}$ in Eq.~(\ref{eq:3.2}) can grow infinitely large as time goes, which is unphysical and just an artifact due to the Taylor expansion.  One may naturally introduce a cutoff onto the expanded $\omega_{\bm p}$ (\ref{eq:3.2}), or accordingly to the time $t$, by remembering that the original $\omega_{\bm p}$ before the expansion has a maximum during the first oscillation $\omega_{\bm p}^{\rm cut} = \omega_{\bm p}^{\rm max} \sim \omega_{\bm p} ( \frac{3\pi}{2} m_\phi^{-1} )$ (see Fig.~\ref{fig;omegaeff}).

From the expansion (\ref{eq:3.2}), the parameter $A$ in Sec.~\ref{sec:2} can be identified as $A = \frac{\sqrt{\zeta }Cm_\phi^2}{\pi}$ in the present model (\ref{phsetup}).  Therefore, the number of produced daughter $\psi$ particles can be estimated with the non-perturbative formula (\ref{nex2}) as 
\begin{align}
    n_{\bm k}^\psi
               = \lambda^2 \pi \left( \frac{\pi}{2} \right)^{\frac{3}{2}} \int \frac{{\rm d}^3{\bm p}}{(2\pi)^3}\frac{\exp\left(-\frac{3\pi^2}{4} \frac{\mu_{\bm p}^2}{\sqrt{\zeta }Cm_\phi^2}\right)}{\left(\sqrt{\zeta }C m_\phi^2 \right)^{\frac32}\Omega_{\bm k}\Omega_{\bm k+\bm p}}\left|D_{\frac{{\rm i}\pi}{2} \frac{\mu_{\bm p}^2}{\sqrt{\zeta }Cm_\phi^2}-\frac12}\left({\rm e}^{{\rm i}\pi/4}\sqrt{\frac{2\pi}{\sqrt{\zeta}Cm_\phi^2}}(\Omega_{\bm k}+\Omega_{\bm k+\bm p})\right)\right|^2 \;, \label{eq:3.3}
\end{align}
where $\mu_{\bm p}^2 = m_{\chi}^2 + {\bm p}^2$.  Note that this formula is valid for both of the kinematically-allowed and -forbidden regimes.  We discuss implications of the production-number formula (\ref{eq:3.3}) to the preheating in each parameter regime.

\subsection{Kinematically-allowed regime: instant preheating and heavy daughter particle production}

We discuss the particle production in the kinematically-allowed regime such that $\omega_{\bm p}^{\rm min}<\Omega_{\bm k}+\Omega_{\bm k+\bm p}<\omega^{\rm max}_{\bm p}$ (see Fig.~\ref{fig;omegaeff}), for which the kinematic condition $\omega_{\bm p} = \Omega_{\bm k} + \Omega_{\bm k+\bm p}$ is satisfied at some time during the first oscillation and the perturbative particle production occurs.  For the condition $\omega_{\bm p}^{\rm min}<\Omega_{\bm k}+\Omega_{\bm k+\bm p}$ to hold, $2m_\psi > M_{\chi}^{\rm min} = m_\chi$ must hold, i.e., the kinematically-allowed particle production is a mechanism to produce heavy daughter $\psi$ particles via decay of relatively light parent $\chi$ particles.  Note that this is the parameter regime that the original work on instant preheating~\cite{Felder:1998vq} considered\footnote{In the original work on the instant preheating~\cite{Felder:1998vq}, the authors considered a case that daughter particles are fermions which couple to a parent scalar field, while our daughter particle $\psi$ is scalar.  It is a straightforward task to extend the present formalism to the fermionic case, which we leave for future work.  }.

From the asymptotic form (\ref{asym}), the particle-number density in the kinematically-allowed regime can be approximated as
\begin{align}
    n_{\bm k}^\psi
        \sim \lambda^2 \frac{\pi^2}{4\sqrt{\zeta }Cm_\phi^2} \int \frac{{\rm d}^3{\bm p}}{(2\pi)^3}\frac{\exp\left(-\pi^2 \frac{\mu_{\bm p}^2}{\sqrt{\zeta }Cm_\phi^2}\right)}{\Omega_{\bm k}\Omega_{\bm k+\bm p}\sqrt{(\Omega_{\bm k}+\Omega_{\bm k+\bm p})^2-\mu_{\bm p}^2}} \;, \label{eq:3.5}
\end{align}
where the error term ${\mathcal O}\left( \exp\left( -\pi \frac{ \mu^2_{\bm p}}{A} \right) \right) = {\mathcal O}\left(\exp\left(-\pi^2\frac{\mu_{\bm p}^2}{\sqrt{\zeta }Cm_\phi^2}\right)\right)$ may be dropped as $\chi$ is assumed to be heavy $m^2_\chi \gtrsim \sqrt{\zeta} C m^2_\phi$.  Equation~(\ref{eq:3.5}) is consistent with the phenomenological perturbative formula (\ref{nLCFA}).  One may further simplify Eq.~(\ref{eq:3.5}) by carrying out the ${\bm p}$ integration within the saddle-point method.  The saddle-point condition is $0 = \partial_{\bm p} \left(-\pi^2\frac{\mu_{\bm p}^2}{\sqrt{\zeta }Cm_\phi^2}\right) = {\rm const.} \times {\bm p}$, and therefore $\bm p=0$ is the saddle point.  Carrying out the saddle-point integration, 
\begin{align}
    n_{\bm k}^\psi
       \sim \frac{\lambda^2}{32\pi^{\frac{5}{2}}} \frac{\left( \sqrt{\zeta}C m_\phi^2 \right)^{\frac12} }{\Omega_{\bm k}^2\sqrt{4\Omega_{\bm k}^2-m_\chi^2}} \exp\left(-\pi^2\frac{m_\chi^2}{\sqrt{\zeta }Cm_\phi^2}\right) \; . \label{eq:-3.5}
\end{align}
Notice that there is a cutoff on the momentum ${\bm k}$ set by the kinematically-allowed condition $\Omega_{\bm k}+\Omega_{\bm k+\bm p}<\omega^{\rm max}_{\bm p}$.  As ${\bm p}={\bm 0}$ dominates the particle production, the momentum ${\bm k}$ shall be constrained by 
\begin{align}
    0   >  ( \Omega_{\bm k}+\Omega_{\bm k+\bm p}- \omega^{\rm max}_{\bm p})|_{{\bm p}={\bm 0}} 
        = 2\Omega_{\bm k} - M^{\rm max}_\chi \;, \label{eq:3.6}
\end{align}
with 
\begin{align}
    M^{\rm max}_\chi \sim \sqrt{m_\chi^2 + \zeta \phi^2\left(\frac{3\pi}{2}m_\phi^{-1}\right)} \sim \sqrt{m_\chi^2+\frac{4\zeta C^2m_\phi^2}{9\pi^2}} \sim \frac{2\sqrt{\zeta} C m_\phi}{3\pi}   
\end{align}
being the maximal mass of $\chi$ during the first oscillation (see Fig.~\ref{fig;omegaeff}).  Solving Eq.~(\ref{eq:3.6}) yields
\begin{align}
    |{\bm k}| < \sqrt{\frac{(M^{\rm max}_\chi)^2}{4} - m_\psi^2 } = k_{\rm max} \;,
\end{align}
which sets the cutoff momentum for the kinematically-allowed particle production.  We comment that the momentum cut-off here is also relevant to the applicability of our formula~\eqref{nex2} [see also discussions below Eq.~(\ref{nex2})]. Although we took the time integration over $[-\infty,\infty]$ to derive Eq.~\eqref{nex2}, the approximation~\eqref{eq:3.2} is valid only within a finite time scale $\mathcal{O}(1/\omega) =\mathcal{O}(1/m_\phi)$.  On the other hand, the time at which the kinematic condition is satisfied is
\begin{align}
    t-\frac{\pi}{ m_\phi} = \frac{\pi}{m_\phi}\sqrt{\left(\frac{\Omega_{\bm k}+\Omega_{\bm k+\bm p}}{\sqrt\zeta Cm_\phi}\right)^2-\left(\frac{\mu_{\bm p}}{\sqrt\zeta Cm_\phi}\right)^2}.
\end{align}
Therefore, the time scale of the decay process is sufficiently smaller than $\mathcal{O}(1/m_\phi)$ provided $\sqrt{\left(\frac{\Omega_{\bm k}+\Omega_{\bm k+\bm p}}{\sqrt\zeta Cm_\phi}\right)^2-\left(\frac{\mu_{\bm p}}{\sqrt\zeta Cm_\phi}\right)^2}\ll1$. In such a case, extending the time integration to an infinite range would be a good approximation, namely Eq.~\eqref{nex2} is valid. On the other hand, when $\left(\frac{\Omega_{\bm k}+\Omega_{\bm k+\bm p}}{\sqrt\zeta Cm_\phi}\right)^2\sim 1$, which is the case roughly for $k\sim k_{\rm max}$, taking the time integration over an infinite range would not be valid. Furthermore, if the kinematic condition is satisfied near the maximal value of $M_\chi(t)$, the actual form of the effective frequency $\omega^2_{\bm p}$ cannot be approximated by the quadratic function of $t$, and the detailed functional form is necessary to estimate the correct value of the produced particle number density. Therefore, evaluating the number density or the energy density with the cut-off $k=k_{\rm max}$ may overestimate/underestimate their actual values, but we expect the error does not change the order of the estimation. A more precise evaluation may be possible e.g. by developing semi-classical methods, which is beyond the scope of this work.\footnote{If we evaluate $\psi$-particle number density with WKB mode functions, one would find that the kinematic condition corresponds to the saddle point condition of the time integration. Then, we would be able to estimate the possible error of the integration with the saddle point method, which may improve our estimation.}

Let us discuss the implications of the particle production in the kinematically-allowed regime (\ref{eq:-3.5}) by estimating the energy density of the produced daughter particles $\rho_\psi$ with the number-density formula (\ref{eq:-3.5}): 
\begin{align}
    \rho_\psi(t) 
        &= a^{-3}(t)\int^{|\bm k|=k_{\rm max}} \frac{{\rm d}^3{\bm k}}{(2\pi)^3} \Omega_{\bm k} n_{\bm k}^\psi \nonumber\\
        &\sim \frac{\lambda^2}{128\pi^{\frac{9}{2}}} m_\psi \left( \frac{m_\phi t}{\pi} \right)^{-2} \left(\sqrt{\zeta} C m_\phi^2 \right)^{\frac{1}{2}} \nonumber\\
            &\quad \times \exp\left(-\pi^2 \frac{m_\chi^2}{\sqrt{\zeta }Cm_\phi^2}\right) \left( \sqrt{ \left(\frac{M^{\rm max}_\chi}{2m_\psi}\right)^2 -1} - \arctan \sqrt{ \left(\frac{M^{\rm max}_\chi}{2m_\psi}\right)^2 -1} \right) \nonumber\\
        &\sim \left\{ \begin{array}{ll} \displaystyle \frac{\lambda^2}{384\pi^{\frac{7}{2}}}  \frac{\left(\sqrt{\zeta} C m_\phi^2 \right)^{\frac{3}{2}}}{m_\phi ( m_\phi t )^{2}} \exp\left(-\pi^2 \frac{m_\chi^2}{\sqrt{\zeta }Cm_\phi^2}\right) & {\rm for}\ M_\chi^{\rm max} \gtrsim 2m_\psi \\ \displaystyle  \frac{\lambda^2}{96\sqrt{2}\pi^{\frac{5}{2}}} \frac{ m_\psi \left(\sqrt{\zeta} C m_\phi^2 \right)^{\frac{1}{2}} }{( m_\phi t )^{2}} \exp\left(-\pi^2 \frac{m_\chi^2}{\sqrt{\zeta }Cm_\phi^2}\right) \left( \left(\frac{1}{3\pi} \frac{\sqrt{\zeta}Cm_\phi}{m_\psi}\right) -1 \right)^{\frac{3}{2}} & {\rm for}\ M_\chi^{\rm max} \sim 2m_\psi \end{array} \right. \; .\label{eq:3.9}
\end{align}
where we used $\sqrt{4\Omega_{\bm k}^2-m^2_\chi} \sim 2\Omega_{\bm k}$, as $2m_\psi > m_\chi$ in the kinematically-allowed region, to get the second line.

We compare the energy density $\rho_\psi$ (\ref{eq:3.9}) with those of the parent particles $\rho_\chi$ and the inflaton $\rho_\phi$ at time $t \sim \frac{3\pi}{2} m_\phi^{-1}$, where the parent particles may have the largest energy density and hence it is kinematically reasonable to assume that a significant portion of the daughter particles is produced.  The energy density of the parent particles $\rho_\chi$ can be estimated as
\begin{align}
    \rho_\chi \left( \frac{3\pi}{2} m_\phi^{-1} \right)
       &= a^{-3} \left( \frac{3\pi}{2} m_\phi^
        {-1} \right) \int \frac{{\rm d}^3{\bm p}}{(2\pi)^3} \omega_{\bm p}^{\rm max} n_{\bm p}^{\chi}  \nonumber\\
       &\sim \frac{1}{27\pi^{\frac{11}{2}}}  \frac{\left(\sqrt{\zeta} C m_\phi^2 \right)^{\frac{5}{2}}}{m_\phi}  \exp\left( -\pi^2\frac{m_\chi^2}{\sqrt{\zeta }Cm_\phi^2} \right) \; ,
\end{align}
where we used $\omega_{\bm p}^{\rm max} \sim \omega_{\bm p} ( \frac{3\pi}{2} m_\phi^{-1} ) 
\sim \frac{2\sqrt{\zeta} C m_\phi}{3\pi}$ and $n_{\bm p}^{\chi} = \exp\left( -\pi^2\frac{m_\chi^2 + {\bm p}^2 }{\sqrt{\zeta }Cm_\phi^2} \right)$ [see Eq.~(\ref{eq::2.32})].  Comparing $\rho_\chi$ with $\rho_\psi$, we find
\begin{align}
   \left. \frac{\rho_{\psi}}{\rho_\chi} \right|_{t=\frac{3\pi}{2} m_\phi^{-1}}
        &\sim  \lambda^2 \frac{3\pi}{32} \frac{ m_\psi m_\phi }{ \left(\sqrt{\zeta} C m_\phi^2 \right)^{2} } \left( \sqrt{ \left(\frac{M^{\rm max}_\chi}{2m_\psi}\right)^2 -1} - \arctan \sqrt{ \left(\frac{M^{\rm max}_\chi}{2m_\psi}\right)^2 -1} \right) \nonumber\\
        &\sim \left\{ \begin{array}{ll} \displaystyle \frac{\lambda^2}{32} \frac{1}{\sqrt{\zeta} C m_\phi^2 } & {\rm for}\ M_\chi^{\rm max} \gtrsim 2m_\psi \\ \displaystyle \frac{\pi \lambda^2}{8\sqrt{2}} \frac{m_\psi m_\phi }{ \left(\sqrt{\zeta} C m_\phi^2 \right)^{2} } \left( \left(\frac{1}{3\pi} \frac{\sqrt{\zeta}Cm_\phi}{m_\psi}\right) -1 \right)^{\frac{3}{2}}    & {\rm for}\ M_\chi^{\rm max} \sim 2m_\psi \end{array} \right. \; . \label{eq-3.11}
\end{align}
The ratio $\rho_{\psi}/\rho_\chi$ does not have any non-perturbative dependencies in the inflaton field $\propto \sqrt{\zeta}C m_\phi^2$, which is the manifestation of that the decay process is driven by the purely perturbative mechanism in the kinematically-allowed regime.  The ratio approaches zero with $M_\chi^{\rm max} \to 2m_\psi$, meaning that the daughter-particle production is inefficient at around the kinematic threshold.  A significant amount of daughter particles can be produced in the other regime $M_\chi^{\rm max} \gtrsim 2m_\psi$.  As an example of an order estimate, let us take $\zeta=1.0 \times 10^{-1}$, $m_\phi = 1.0\times 10^{13}\;{\rm GeV}$, $m_\chi=2.5\times 10^{15}\;{\rm GeV}$, $m_\psi=1.0\times 10^{17}\;{\rm GeV}$, $\lambda = 5.6\times 10^{16}\;{\rm GeV}$ (corresponding to $C = 3.9\times10^{5}$).  The resulting energy ratio is $\rho_\psi/\rho_\chi = 0.9$.  This means that most energy of $\chi$ can be converted to that of $\psi$ immediately within the first inflaton oscillation if the parameters are properly chosen, as proposed in Ref.~\cite{Felder:1998vq}.  Note that the energy ratio $\rho_\psi/\rho_\chi$ (\ref{eq-3.11}) is a monotonically increasing function of $\lambda$, which seems to break the energy conservation $\rho_\psi/\rho_\chi > 1$ eventually.  For such large values of $\lambda$, the estimate (\ref{eq-3.11}) is inapplicable because it is based on our lowest-order treatment in $\lambda$ [see Fig.~\ref{fig1} or the resulting formula (\ref{nex2})] and one needs to include higher-order $\lambda$ corrections.  The estimate (\ref{eq-3.11}) implies that the lowest-order treatment in $\lambda$ is justified for $4\sqrt{2}\left( \sqrt{\zeta}Cm_\phi^2 \right)^{1/2}\gtrsim\lambda$ when $M_\chi^{\rm max}\gtrsim 2m_\psi$.

We turn to compare $\rho_\psi$ (\ref{eq:3.9}) with the inflaton energy density $\rho_\phi$.  At $t \sim \frac{3\pi}{2} m_\phi^{-1}$, $\rho_\phi$ can be estimated as
\begin{align}
    \rho_\phi \left( \frac{3\pi}{2} m_\phi^{-1} \right)
        = \left. \frac{1}{2} \left( \dot{\phi}^2 + m_\phi^2 \phi^2 \right) \right|_{t=\frac{3\pi}{2} m_\phi^{-1}}
        \sim \frac{2}{9\pi^2} C^2m_\phi^4 \;.
\end{align}
Therefore, the energy fraction released as $\psi$ particles can be estimated as 
\begin{align}
    \left. \frac{\rho_{\psi}}{\rho_\phi} \right|_{t=\frac{3\pi}{2} m_\phi^{-1} }
        &\sim \frac{\zeta \lambda^2}{64\pi^{\frac{5}{2}}}    \frac{m_\psi }{\left(\sqrt{\zeta} C m_\phi^2 \right)^{\frac{3}{2}}}   \nonumber\\
            &\quad \times \exp\left(-\pi^2 \frac{m_\chi^2}{\sqrt{\zeta }Cm_\phi^2}\right) \left( \sqrt{ \left(\frac{M^{\rm max}_\chi}{2m_\psi}\right)^2 -1} - \arctan \sqrt{ \left(\frac{M^{\rm max}_\chi}{2m_\psi}\right)^2 -1} \right) \nonumber\\
        &\sim \left\{ \begin{array}{ll} \displaystyle \frac{
        \zeta \lambda^2}{192
        \pi^{\frac{7}{2}}} \frac{ 1}{m_\phi \left(\sqrt{\zeta} C m_\phi^2 \right)^{\frac{1}{2}}}  \exp\left(-\pi^2 \frac{m_\chi^2}{\sqrt{\zeta }Cm_\phi^2}\right) & {\rm for}\ M_\chi^{\rm max} \gtrsim 2m_\psi \\ \displaystyle \frac{
        \zeta \lambda^2}{48\sqrt{2}\pi^{\frac{5}{2}}}  \frac{m_\psi  }{ \left(\sqrt{\zeta} C m_\phi^2 \right)^{\frac{3}{2}} }\left( \left(\frac{1}{3\pi} \frac{\sqrt{\zeta}Cm_\phi}{m_\psi}\right) -1 \right)^{\frac{3}{2}} \exp\left(-\pi^2 \frac{m_\chi^2}{\sqrt{\zeta }Cm_\phi^2}\right)  & {\rm for}\ M_\chi^{\rm max} \sim 2m_\psi \end{array} \right. \; . \label{eq--3.13}
\end{align}
Note that the energy ratio of the parent $\chi$ particles (before decaying into $\psi$) to the inflaton, $\rho_\chi/\rho_\phi$, can be estimated as 
\begin{align}
    \left. \frac{\rho_\chi}{\rho_\phi} \right|_{t=\frac{3\pi}{2} m_\phi^{-1} }
        \sim \frac{\zeta^{\frac54}\sqrt{C}}{6\pi^{\frac72}}\exp\left(-\pi^2 \frac{m_\chi^2}{\sqrt{\zeta }Cm_\phi^2}\right) \;.
\end{align}
For the same parameter set considered below Eq.~(\ref{eq-3.11}), the energy fraction released as $\psi$ is $\rho_\psi/\rho_\phi = 6.6\times 10^{-4} $, while that as $\chi$ is $\rho_\chi/\rho_\phi = 7.3\times 10^{-4}$ (remind that $\rho_\psi/\rho_\chi = 0.9$).  The closeness of the values, $\rho_\psi/\rho_\phi$ and $\rho_\chi/\rho_\phi$, means that the exponential suppression of the parent $\chi$ particle production $n^\chi_{\bm p} \propto \exp \left( -\pi^2 \frac{m^2_\chi}{\sqrt{\zeta}Cm^2_\phi}\right)$ determines the typical magnitude of the ratios and the other factors only give subleading contributions, as the decay is perturbative and hence there is no non-perturbative correction from the decay probability.  The smallness of the energy fraction $\rho_\psi/\rho_\phi \sim 0.07\;\%$ validates neglecting the backreaction of the particle production to the inflaton $\phi$.  Even though the fraction right after the production $t \sim \frac{3\pi}{2} m_\phi^{-1}$ is small, as discussed in Ref.~\cite{Felder:1998vq,Felder:1999pv}, the daughter particle $\psi$ may be able to dominate the Universe if the inflaton energy decays faster than $a^{-3}(t)$, which may happen when the inflaton potential differs from $\frac12 m_\phi^2\phi^2$.  Such a modification would change some details of our result but not qualitatively.  Note that taking lighter $m_\chi$ increases the energy fraction, meaning that the energy of the inflaton $\phi$ would be converted efficiently to the heavy daughter $\psi$ particles and thus $\psi$ may dominate the Universe quickly.  In such a case, however, one needs to include the backreaction (i.e., the inflaton field $\phi$ should be treated dynamically rather than as a background) to make a proper estimate, which is beyond our framework.

\subsection{Kinematically-forbidden regime: light daughter particle production} \label{sec3.2}

We discuss the kinematically-forbidden particle production, i.e., the production of light daughter $\psi$ particles from heavy parent $\chi$ particles.  Namely, we consider a case $m_{\chi} \gg 2m_\psi$, for which the kinematic condition $\omega_{\bm p} = \Omega_{\bm k} + \Omega_{\bm k+\bm p}$ is unsatisfied during the oscillation for not-large values of ${\bm k}$ because $\Omega_{\bm k}+\Omega_{\bm k+\bm p} \sim 2m_\psi \ll m_\chi < \omega_{\bm p}$ [see 
also the discussions below Eq.~\eqref{asym} and 
Eq.~\eqref{eq:3.019} for that the kinematic condition can be satisfied when $|{\bm k}|$ becomes large $|{\bm k}|/m_\psi \gtrsim m_\chi/(2m_\psi)$].  Thus, the naive perturbative estimation fails completely.  We shall estimate the daughter particle number and its energy density in the kinematically-forbidden regime and discuss its implication to cosmology.

The asymptotic form of the particle-number density in the kinematically-forbidden regime reads [see Eq.~(\ref{asym})] 
\begin{align}
    n_{\bm k}^\psi
    &\sim \lambda^2 \frac{\pi^2}{4\sqrt{\zeta}Cm_\phi^2} \int\frac{{\rm d}^3\bm p}{(2\pi)^3}\frac{ \exp\left[-\pi^2\frac{\mu_{\bm p}^2}{\sqrt{\zeta}Cm_\phi^2}\left( 1 - F\left( \frac{\Omega_{\bm k}+\Omega_{\bm k+\bm p}}{\mu_{\bm p}} \right) \right) \right] }{ \Omega_{\bm k}\Omega_{\bm k+\bm p}\sqrt{\mu_{\bm p}^2-(\Omega_{\bm p}+\Omega_{\bm k+\bm p})^2}} \;,
\end{align}
where we dropped $\exp\left(-2 \pi^2 \frac{\mu_{\bm p}^2}{\sqrt{\zeta}C m_\phi^2} \right)$ contributions and used $F \in {\mathbb R}$ for $ \mu_{\bm p}  >  \Omega_{\bm k}+\Omega_{\bm k+\bm p}$.  We apply the saddle point method to the $\bm p$ integration.  The saddle condition is
\begin{align}
    {\bm 0}  
        &= \partial_{\bm p} \left[  -\pi^2\frac{\mu_{\bm p}^2}{\sqrt{\zeta}Cm_\phi^2}\left( 1 - F\left( \frac{\Omega_{\bm k}+\Omega_{\bm k+\bm p}}{\mu_{\bm p}} \right) \right)  \right] \nonumber\\
        &= -2\pi\frac{\mu_{\bm p}}{\sqrt{\zeta}C m_\phi^2} \frac{\bm k+\bm p}{\Omega_{\bm k+\bm p}} \left[ 1 + {\mathcal O}\left( \frac{\Omega_{\bm k}+\Omega_{\bm k+\bm p}}{\mu_{\bm p}} \right) \right] \; ,
\end{align}
and thus the saddle point can be identified as $\bm p=-\bm k$ after neglecting ${\mathcal O}\left( \frac{\Omega_{\bm k}+\Omega_{\bm k+\bm p}}{\mu_{\bm p}} \right)$.  Carrying out the saddle-point integration, 
\begin{align}
    n_{\bm k}^\psi
    &\sim  \frac{\lambda^2}{32\pi} \left( \sqrt{\zeta}Cm_\phi^2 \right)^{\frac{1}{2}} \sqrt{\frac{m_\psi}{\mu_{\bm k}}} \frac{ \exp\left[-\pi^2\frac{\mu_{\bm k}^2}{\sqrt{\zeta}Cm_\phi^2}\left( 1 - F\left( \frac{\Omega_{\bm k}+m_\psi}{\mu_{\bm k}} \right) \right) \right] }{ \mu_{\bm k} \Omega_{\bm k} \sqrt{\mu_{\bm k}^2-(\Omega_{\bm k}+m_\psi)^2}}  \; .  \label{eq::3.17}
\end{align}
Note that there is a momentum cutoff on $|{\bm k}|$ set by the kinematically-forbidden condition $\Omega_{\bm k}+\Omega_{\bm k+\bm p}\leq \mu_{\bm p}$, or $\Omega_{\bm k}+m_\psi \leq \mu_{\bm k}$ as ${\bm p} \sim -{\bm k}$ gives the dominant contribution.  Thus, the cutoff momentum $k_{\rm max}$ can be estimated as
\begin{align}
   |{\bm k}| 
    < m_\chi\sqrt{\left(\frac{m_\chi}{2m_\psi}\right)^2-1} 
    = k_{\rm max} \; , \label{eq:3.019}
\end{align}
which is effectively infinite $k_{\rm max} \to \infty$ in the limit of $m_\chi/m_\psi \to \infty$.

Using the number-density formula (\ref{eq::3.17}), we can estimate the energy density of $\psi$ produced: 
\begin{align}
   \rho_\psi(t) 
        &= a^{-3}(t) \int \frac{{\rm d}^3{\bm k}}{(2\pi)^3} \Omega_{\bm k} n^\psi_{\bm k} \nonumber\\
        &\sim \frac{\lambda^2}{32 \pi} \left( \frac{m_\phi t}{\pi} \right)^{-2} \left( \sqrt{\zeta}Cm_\phi^2 \right)^{\frac{1}{2}} \int \frac{{\rm d}^3{\bm k}}{(2\pi)^3} \sqrt{\frac{m_\psi}{\mu_{\bm k}}} \frac{ \exp\left[-\pi^2\frac{\mu_{\bm k}^2}{\sqrt{\zeta}Cm_\phi^2}\left( 1 - F\left( \frac{\Omega_{\bm k}+m_\psi}{\mu_{\bm k}} \right) \right) \right] }{ \mu_{\bm k} \sqrt{\mu_{\bm k}^2-(\Omega_{\bm k}+m_\psi)^2}}  \; .
\end{align}
As the exponent is a function of ${\bm k}^2$, the saddle point of the ${\bm k}$ integration can be identified as $\bm k=0$.  Thus, carrying out the saddle-point integration, 
\begin{align}
   \rho_\psi(t) 
        &\sim \frac{\lambda^2}{(4\pi)^4} \left( \frac{m_\phi t}{\pi} \right)^{-2} \left( \frac{m_\psi}{m_\chi} \right)^2 \frac{ \left( \sqrt{\zeta}Cm_\phi^2 \right)^2 }{ m_\chi^2} \exp\left( -\frac{\pi^2}{2} \frac{m_\chi^2}{\sqrt{\zeta}Cm_\phi^2} \right)  \; ,
\end{align}
where we have used $ m_\chi \gg 2m_\psi$ and extended the $|{\bm k}|$ integration range from $[0,k_{\rm max}]$ to $[0,\infty]$.

Let us compare the energy density of $\psi$ with those of $\chi$ and $\phi$.  Note that we shall not do this comparison at $t\sim \frac{3\pi}{2} m_\phi^{-1}$, unlike in the kinematically-allowed case.  The reason is the following.  It was reasonable in the kinematically-allowed regime to assume that the decay $\ket{\chi}\to\ket{\psi\psi}$ occurs at around $t \sim \frac{3\pi}{2} m_{\phi}^{-1}$, where the parent $\chi$ particle has the largest energy and hence has the largest number of kinematical channels to decay.  However, in the kinematically-forbidden regime, such a kinematic argument may not hold.  In fact, $\psi$ can be produced, even if the kinematics of the decay $\ket{\chi}\to\ket{\psi\psi}$ looks unsatisfied, the background inflaton $\phi$ field can compensate the kinematics through multiply scattering with the particles.  This implies that the $\psi$ production does not necessarily occur at $t \sim \frac{3\pi}{2} m_{\phi}^{-1}$ but can occur at {\it anytime} if $\phi$ is sufficiently strong.  Thus, in principle, one cannot determine the time when the $\psi$ production occurs the most in the kinematically-forbidden regime.\footnote{It may be possible to estimate the production time if we perform time-integration e.g. by the saddle point method.  We leave the question of the production time for future work.}  For this reason, we do not do the comparison at $t\sim \frac{3\pi}{2} m_\phi^{-1}$ but rather take it to be the asymptotic future $m_\phi t \gg 1$, where the ratios of the energy densities become invariant under the time evolution [provided that the matter particles are sufficiently heavy so that they are non-relativistic and then the energy densities scale as $a^{-3} \propto t^{-2}$ at the asymptotic times] and hence we can make an unambiguous comparison insensitive to the production time.  Also, considering the asymptotic time is physically motivated, as it is related to the current observation of the Universe (provided that the backreaction is negligible).

The energy densities of $\chi$ and $\phi$ for sufficiently large time $t$ are given, respectively, by
\begin{align}
   \rho_\chi(t) 
        &= a^{-3}(t) \int \frac{{\rm d}^3{\bm p}}{(2\pi)^3} \omega_{\bm p}(t) n^\chi_{\bm p} \nonumber\\
        &\sim \frac{1}{8\pi^{\frac{9}{2}}}  \left( \frac{m_\phi t}{\pi} \right)^{-2} m_\chi \left(\sqrt{\zeta} C m_\phi^2 \right)^{\frac{3}{2}} \exp\left( -\pi^2\frac{m_\chi^2}{\sqrt{\zeta }Cm_\phi^2} \right) \; ,
\end{align}
where we used $M_\chi(t) \to m_\chi$ (i.e., the inflaton-dependent mass becomes negligible as the inflaton amplitude decays at the asymptotic future), and
\begin{align}
    \rho_\phi(t) 
        = \frac{1}{2} \left( \dot{\phi}^2 + m_\phi^2 \phi^2 \right) 
        \sim \frac{C^2m_\phi^2 }{2t^2} \; .
\end{align}
Taking the ratios of $\rho_\psi$ to $\rho_\chi$ and to $\rho_\phi$, we have 
\begin{align}
    \frac{\rho_\psi}{\rho_\chi}
        &\sim  \frac{1}{32\sqrt{\pi}} \left(\frac{ \sqrt{\zeta}Cm_\phi^2}{\pi^2m_\chi^2} \right)^{\frac{1}{2}}\left(\frac{\lambda}{m_\chi}\right)^2 \left( \frac{m_\psi}{m_\chi} \right)^2 \exp\left( +\frac{\pi^2}{2} \frac{m_\chi^2}{\sqrt{\zeta}Cm_\phi^2} \right)  \; , \label{eq::3.23}
\end{align}
and\begin{align}
    \frac{\rho_\psi}{\rho_\phi}
        &\sim \frac{\zeta}{128 \pi^2} \left(\frac{\lambda}{ m_\chi}\right)^2\left( \frac{m_\psi}{m_\chi} \right)^2 \exp\left( -\frac{\pi^2}{2} \frac{m_\chi^2}{\sqrt{\zeta}Cm_\phi^2} \right)\; .\label{eq::3.24} 
\end{align}

What is crucially important here is that the ratio $\rho_\psi/\rho_\chi$ (\ref{eq::3.23}) has an exponential factor with a positive exponent.  Such a non-perturbative effect was absent in the kinematically-allowed case (\ref{eq-3.11}), where the decay process is purely perturbative, which neglects the multiple scattering effects by the inflaton field.  This exponential enhancement becomes important when $\chi$ becomes very heavy (or $\sqrt{\zeta} C m_\phi^2$ small) and can overcome the strong suppression factor of $( m_\psi/m_\chi )^2 \ll 1$.  The new particle production process is not confined to the perturbative constraint $\rho_\psi/\rho_\chi<1$ as the process is free from the kinematic constraint, namely the instantaneous energy conservation due to the energy supply by the inflaton field.

Although the energy ratio between $\psi$ and $\chi$ can be exponentially large, this does not mean that $\psi$ particles immediately dominate the Universe.  In fact, the energy ratio between $\psi$ and $\phi$ (\ref{eq::3.24}) has the exponential factor with a negative exponent and is always suppressed strongly by the mass $m_\chi$.  That is, the inflaton energy released as $\psi$ particles must be small and does not exceed unity, as was the case in the kinematically-allowed case (\ref{eq--3.13}).  Nonetheless, the exponential suppression in the kinematically-forbidden regime (\ref{eq::3.24}) is halved compared to the kinematically-allowed one (\ref{eq--3.13}).  This is the manifestation of the non-perturbative effect that produces $|\psi\psi\rangle$ through multiple scattering between the inflaton field and the particles.  

A salient feature of the kinematically-forbidden process appears when the coupling between $\chi$ and the inflaton is small or the inflaton amplitude $C$ is not large. A naive perturbative estimation leads to a negligibly small amount of $\psi$ production in such a case, but due to the halving of the exponent, a sizable number of $\psi$ can be produced.  Let us show an illustrating example: Taking $m_\phi=1.0\times 10^{13}\;{\rm GeV}$, $m_\chi=\lambda = 1.3\times 10^{15}\;{\rm GeV}$, $m_\psi=5.0\times 10^{13}\;{\rm GeV}$, $\zeta=1.0\times 10^{-4}$, we find $\exp\left( +\frac{\pi^2}{2} \frac{m_\chi^2}{\sqrt{\zeta}Cm_\phi^2} \right) = 4.9\times 10^{8}$, which results in $\rho_\psi/\rho_\chi = 2.0 \times 10^{3}$ and $\rho_\psi/\rho_\phi = 2.6\times 10^{-19}$.  Note that although the energy-density ratio $\rho_\psi/\rho_\phi$ seems negligibly small, this amount of $\rho_\psi$ may become compatible with today's dark matter density if the reheating temperature is $\mathcal{O}(10^9\;{\rm GeV})$~\cite{Chung:1998zb,Chung:1998bt}.\footnote{Since $\chi$ is much heavier than $\psi$, $m_\chi \gg m_\psi$, it is kinematically forbidden for $\psi$ to decay perturbatively into $\chi$.  Therefore, the heavy $\psi$ can be a stable dark matter candidate.  The loop correction through $\zeta\phi^2\chi^2$ and $\lambda\chi\psi^2$ may lead to a decay channel $\ket{ \psi\psi} \to \ket{ \phi\phi}$ but it would not be large.  } In this example, the coupling between the inflaton and the parent particle is very tiny, and the production of $\chi$ is negligibly small.  Nevertheless, what is surprising here is that the daughter particle $\psi$ indirectly coupled to the inflaton is more efficiently produced than $\chi$, and it can produce a sizable amount of energy density, which may become important in the subsequent evolution of the Universe.

\section{Summary and outlook}\label{sec4}

We have provided a fully quantum-field theoretic formulation of the particle production for interacting particles dressed by a background inflaton field in the preheating phase based on the Furry-picture perturbation theory.  We have obtained the general formula for the number of parent and daughter particles produced from the vacuum (\ref{eq:2.21}) and the compact expression (\ref{nex2}) for sufficiently slow inflaton field.  We have shown that the number formula (\ref{nex2}) agrees with the phenomenological perturbative treatment [see Eqs.~(\ref{eq::2.35}) and (\ref{nLCFA})] in the kinematically-allowed parameter regime such that it is kinematically possible for the parent particles to decay into the daughter particles via the perturbative mechanism.  Meanwhile, we have found a novel particle production mechanism, dubbed as the kinematically-forbidden particle production, which occurs outside of the kinematically-allowed regime of the perturbative decay and cannot be captured by the phenomenological perturbative treatment (see Figs.~\ref{fig2}, \ref{fig3}, and \ref{fig4}).  Such a kinematically-forbidden particle production is possible because the particles are non-perturbatively scattered by the inflaton field, which compensates the kinematics to assist the production of the daughter particles.  The behavior of the naive ``decay probability'' in the kinematically-forbidden regime~\eqref{Pforbidden} implies that the daughter $\psi$-particle production is not due to the perturbative decay process $|\chi\rangle\to|\psi\psi\rangle$; otherwise the unitarity is violated.  The kinematically-forbidden process occurs due to a non-perturbative-scattering effect with the coherent background field.  The kinematically-forbidden process overwhelms the kinematically-allowed one in a certain parameter regime such as light daughter particles.  In such a parameter regime, the kinematically-forbidden process would play an important role in preheating, and hence should be taken into account for a complete description of the history of the Universe.  

As an outlook, it would be important to consider in more detail the impacts of the kinematically-forbidden process to cosmology. For example, super-hearvy particles are ubiquitous in UV theories such as superstring or supergravity.  Super-heavy fields are supposed to be decoupled from the low-energy dynamics as their production would be exponentially suppressed even at the very beginning of the preheating, and one naively integrates them out.  Our result indicates that a sizable amount of light particles coupled to super-heavy parent particles can be produced via the kinematically-forbidden production, even though the  production of super-heavy particles themselves is negligible, which would never be expected within the naive perturbation theory.  If it is the case, one cannot integrate super-heavy fields out naively as it would miss the light particle production.  Thus, the kinematically-forbidden particle production may change the conventional view of cosmology as the low-energy limit of UV theories.   On the other hand, although we have concentrated on the kinematically-forbidden parameter regime such that the parent particle is much heavier than the daughter particle (see Sec.~\ref{sec3.2}) and this is the only kinematically-forbidden regime in the slow-field limit, for which the inflaton field goes infinity as time goes, there in principle exists another kinematically-forbidden regime, e.g., for inflaton fields with finite amplitude, such that the daughter particle is much heavier than the parent particle.  Such a production mechanism could produce super-heavy WIMP dark matter, which is an interesting possibility deserving further investigation (cf. production of super-heavy dark matter in preheating was also investigated in Refs.~\cite{Chung:1998zb,Chung:1998bt}).  Finally, for a description of more realistic cosmological models, we need to generalize our method to particles with spins, which would be a straightforward generalization of our analysis in this work, although some details would change due to, e.g., the quantum statistics.

Despite the generality of our formalism presented in Sec.~\ref{sec:2}, the compact formula (\ref{nex2}) may be a bit unsatisfactory as the background fields are Taylor expanded up to only the lowest order $\sim t$ and hence non-linear effects beyond the linear approximation, which can be important for non-slow fields, cannot be described.  For instance, our treatment cannot be applied when multiple particle production events equally contribute to particle production, for which, e.g., quantum interference between different production events occur and the production number is significantly modified \cite{Hebenstreit:2009km, SHEVCHENKO20101, Dumlu:2010ua, Dumlu:2010vv, Dumlu:2011cc, Dumlu:2011rr, Taya:2020dco,Taya:2021dcz,Hashiba:2021npn,Yamada:2021kqw,Enomoto:2022mti,Enomoto:2022nuj,Hashiba:2022bzi}.  Such a multiple particle production would be relevant in the preheating, e.g., if the parent particle is light so that the exponential suppression in the parent particle-number density is not small enough.  A possible avenue to deal with more general background fields is semi-classical methods.  Indeed, recent developments in the understanding of the WKB method~\cite{Dumlu:2010ua,Dabrowski:2014ica,Dabrowski:2016tsx,Li:2019ves,Enomoto:2020xlf,Taya:2020dco,Taya:2021dcz,Enomoto:2021hfv,Hashiba:2021npn,Yamada:2021kqw,Hashiba:2022bzi} would enable us to address the non-perturbative dressed mode function in more general backgrounds.  The semi-classical approach would also help us to develop more physical understandings of the kinematically-forbidden process (e.g., the production time).  We report a generalization of our approach based on semi-classical methods elsewhere~\cite{inprep}.

\section*{Acknowledgement}
HT is supported by JSPS KAKENHI under grant No. 22K14045 and the RIKEN special postdoctoral researcher program. YY is supported by Waseda University Grant for Special Research Projects (Project number: 2022C-573). The authors also thank the Nonequilibrium working group at RIKEN Interdisciplinary Theoretical and Mathematical Sciences, thanks to which this collaboration has started.

\bibliographystyle{JHEP}
\bibliography{ref.bib}

\end{document}